\documentclass{achemso}
\usepackage{dcolumn}
\usepackage{hyperref}
\usepackage{multirow}
\usepackage{natbib}
\setkeys{acs}{etalmode=truncate,maxauthors=50}
\usepackage[
]{achemso}
\hypersetup{
    colorlinks=true,
    linkcolor=blue,
    filecolor=magenta,      
    urlcolor=blue,
    citecolor=blue,
}
\makeatletter
\g@addto@macro\normalsize{%
  \setlength\abovedisplayskip{12pt}
  \setlength\belowdisplayskip{12pt}
  \setlength\abovedisplayshortskip{12pt}
  \setlength\belowdisplayshortskip{12pt}
}
\makeatother
\usepackage{physics}
\usepackage{epsfig}
\usepackage{graphics}
\usepackage{float}
\usepackage{amsmath}
\usepackage{xcolor,colortbl}
\usepackage[utf8]{inputenc}
\usepackage[T1]{fontenc}
\usepackage{lmodern} 
\usepackage{soul}
\usepackage[version=4]{mhchem}
\usepackage{soul}

\definecolor{Gray}{rgb}{0.72,0.72,0.98}
\definecolor{LightCyan1}{rgb}{0.83,0.83,0.98}
\definecolor{LightCyan2}{rgb}{0.91,0.92,1}

\renewcommand{\[}{\left[}

\newcommand{\Hightlight}[1]{\textcolor{black}{{#1}}}

\title{Photovoltaic efficiency of transition metal dichalcogenides thin films by ab initio excited-state methods}

\author{Enesio Marinho Jr}
 \email{enesio.marinho@unesp.br}
 \affiliation{
 Departamento de Física e Química, Universidade Estadual Paulista (UNESP),\\
Av.\ Brasil, 56, Ilha Solteira, 15385-007 São Paulo, Brazil.}
\affiliation{
 Instituto de Física Teórica, Universidade Estadual Paulista (UNESP),\\
R.\ Dr.\ Bento Teobaldo Ferraz, 271, São Paulo, 01140-070 São Paulo,  Brazil.}

\author{Cesar E. P. Villegas}
\email{cesar.villegas@unesp.br}
\affiliation{ Departamento de Ciencias, Universidad Privada del Norte, Lima 15434, Peru}

\author{Pedro Paulo de Mello Venezuela}
\email{pedrovenezuela@id.uff.br}
\affiliation{Instituto de Física, Universidade Federal Fluminense (UFF),\\ Av.\ Gal.\ Milton Tavares de Souza, s/n, 24210-346 Niterói, Rio de Janeiro, Brazil.}

\author{Alexandre Reily Rocha}
\email{alexandre.reily@unesp.br}
\affiliation{
 Instituto de Física Teórica, Universidade Estadual Paulista (UNESP),\\
R.\ Dr.\ Bento Teobaldo Ferraz, 271, São Paulo, 01140-070 São Paulo,  Brazil.}

\date{\today}

\begin{document}
\maketitle
\begin{abstract}

Transition metal dichalcogenides (TMDCs) have garnered significant interest in optoelectronics, owing to their scalability and thickness-dependent electrical and optical properties.  In particular, thin films of TMDCs could be used in photovoltaic devices. In this work, we employ \emph{ab initio} many-body perturbation theory within $G_0W_0$-BSE approach to accurately compute the optoelectronic properties of thin films of 2H-TMDCs composed of Mo, W, S, and Se. Subsequently, we evaluate their photovoltaic performance including exciton recombination effects, and show this is a key ingredient. We obtain efficiencies of up to 29\% for a 200-nm thick film of \ce{WSe2}, thus providing an upper limit.  We also include other \textcolor{black}{phenomenological} recombination mechanisms that could be present in current samples. This slightly reduces efficiencies, indicating that even with current synthesis technologies, there is still potential for further enhancement of TMDCs' performance in photovoltaic applications.

\end{abstract}
\begin{center}
\end{center}

\section{Introduction}
Layered van der Waals (vdW) materials consist of stacked two-dimensional (2D) sheets interacting via dispersion forces. Graphite is a well-known member of this family that is broadly used in industry as components for electrodes, lubricants, fibers, heat exchangers, and batteries \cite{vdW-Science}. The experimental isolation of its single-layer counterpart, graphene \cite{Novoselov2004} has opened the way for a new era of 2D materials with compelling optoelectronic properties \cite{khan2020,ma2021,villegas2022}. However, the lack of an energy gap in graphene greatly
limits its applicability in semiconductor technologies. On the other hand, the semiconducting properties of transition metal dichalcogenides (TMDCs), with chemical formula \ce{MX2} (M = Mo, W, and others, and X = S, Se, Te) pushed the frontier of vdW materials' potential uses spanning a broad range of fields, including nanoelectronics and nanophotonics, sensing and actuation in the nanoscale, and photovoltaic solar cells \cite{chhowalla2013NatChem,Manzeli2017NatRevMat,perez2020,villegas2015,britnell-science,furchi2018device,baugher2014optoelectronic}.

The TMDCs monolayers present direct bandgaps ranging from ${\sim} 1.4$ to $2.5$ eV \cite{Kim2021PRB}, which is well-suited for solar energy absorption \cite{Aftab23}, particularly as ultra-thin photon absorbers for applications in transparent and flexible solar cells \cite{nassiri2021high,He2022}. \Hightlight{Theoretical predictions indicate that the TMDCs monolayers should present absorption coefficients per unit thickness one to two orders of magnitude greater than those of conventional semiconductors  \cite{Bernardi-NanoLett2013}. This is achieved due to the localized nature of electronic wave functions, leading to van Hove singularities in the density of states. These strong peaks are accountable for the heightened photoresponsivity observed in these 2D materials \cite{britnell2013Science}.} As a consequence, with a thickness of less than 1 nm, in principle, they can absorb an amount of visible light comparable to  15 nm of GaAs or 50 nm of Si \cite{Bernardi-NanoLett2013}, which makes them a promising choice for efficient photovoltaic applications.
However, in practice, the power conversion efficiency (PCE) of single layer TMDC solar cells has typically not exceeded 2\% \cite{li2015ultimate,Cheng14,mao2018magnetron,went2019new,mcvay2020impact,Islam22}. This is mostly due to strong Fermi-level pinning at the metal–semiconductor interface \cite{liu2018approaching}, and high probability of carrier recombination caused by defective surfaces, which leads to a low open-circuit voltage ($V_{\text{oc}}$) \cite{nassiri2021high}. 

In order to overcome those performance issues, some TMDC-based solar cells have been assembled by using few-layered or multi-layered stacks of 2H-TMDCs. Lee et al. \cite{lee2014atomically} have studied the photovoltaic response of graphene-sandwiched \ce{MoS2}/\ce{WSe2} heterostructures with different thicknesses. They reported a  high short-circuit current density of ${\sim}2.2$ A/cm$^2$, measured under laser light ($\lambda=532$ nm). For comparison, the authors also reported the short-circuit current of vertical heterojunctions composed of single atomic layers of \ce{MoS2} ($n$-doped) and \ce{WSe2} (either $p$-doped or ambipolar) with a value of ${\sim}1$ mA/cm$^2$. The external quantum efficiency (EQE) is also dependent on the number of layers. The measured EQEs at
532 nm were 2.4\%, 12\%, and 34\% for monolayer, bilayer, and multilayer p–n junctions. Memaran et al. \cite{memaran2015pronounced} have studied the photovoltaic efficiency of the \ce{MoSe2} crystal
composed of ${\sim}10$ atomic layers transferred onto a flat hBN crystal creating an electrostatic p-n junction. They have reported photovoltaic
efficiencies surpassing 14\% under AM-1.5 spectrum, with
concomitant fill factors approaching 0.7.

\textcolor{black}{More recently, Cho et al.~\cite{Cho18} proposed a transparent thin-ﬁlm photovoltaic cell composed by a heterojunction of \ce{WSe2}/\ce{MoS2} and an indium tin oxide electrode. Their device is highly transparent (${\sim}80\%$) and presents a PCE of ${\sim}$10\%. Also, high-speciﬁc-power ﬂexible transition metal dichalcogenide solar cells have been fabricated by 
employing transparent graphene contacts and \ce{MoO_x} capping for doping, passivation, and anti-reﬂection \cite{Nazif21}. These lead to PCE of 5.1\%
and speciﬁc power of 4.4 Wg$^{-1}$ for ﬂexible \ce{WSe2}  solar cells. }

These observations suggest that junctions composed of few-layered or multilayered TMDCs can yield higher photovoltaic responses compared to monolayers. Nonetheless, while monolayer TMDCs have been extensively studied and well characterized for photovoltaic applications, their bulk counterparts were less so, especially from a microscopic point of view. In this sense, a highly accurate and reliable \emph{ab initio} prediction of the optoelectronic properties and photovoltaic efficiencies of bulk 2H-TMDCs is crucial. 

Herein, we employ \emph{ab initio} many-body perturbation theory within the $GW$-BSE scheme to investigate the optoelectronic properties of bulk 2H-TMDCs and then address their theoretical photovoltaic efficiencies. We analyze the power-conversion efficiency of \ce{MX_2} TMDCs with M = Mo, W and X = S, Se, modeling the inclusion of nonradiative recombination losses with exciton eigenvalues computed by BSE, in an approach we denote as SLME-X which has already been proposed by Ozório and co-workers \cite{ozorio2022theoretical}, and compared those results with another modeling in which we estimate the fraction of radiative recombination processes using reported values obtained by photoluminescence quantum yield experiments (SLME-PLQY). Applying the SLME-X approach, the photovoltaic efficiencies of the TMDCs consistently fell within the range of 23\% to 29\%, reaching efficiencies analogous to those obtained using the Shockley-Queisser formalism. On the other hand, when we employ the SLME-PLQY framework, the power conversion efficiencies tend to be reduced, ranging from 19\% to 24\%.

\section{\label{sec:method}Computational Methods}
Ground-state atomic structures were obtained by first-principles calculations based on density-functional theory (DFT) \cite{HohenbergKohn1964, KohnSham1965} as implemented in the \textsc{q}uantum \textsc{espresso} package \cite{qe}. The Perdew-Burke-Ernzerhof generalized-gradient approximation \cite{pbe} was employed to describe the exchange-correlation functional. Fully relativistic 
optimized norm-conserving Vanderbilt pseudopotentials (ONCVPSP) \cite{ONCVPSP}, considering semicore $s$ and $p$ states as valence electrons, are used to describe the electron–ion interaction. A kinetic energy cutoff of $84$ Ry is set to expand the Kohn-Sham orbitals. The charge density is obtained in a $\Gamma$-centered Monkhorst-Pack \emph{k}-point sampling of $8 {\times} 8 {\times} 4$. The structures are fully relaxed including van der Waals corrections within the semi-empirical dispersion scheme (PBE-D2) as proposed by Grimme \cite{pbe-d2}, and also including spin-orbit coupling.

The many-body perturbation theory (MBPT) framework within the $GW$ approximation \cite{hedin1965} has been applied to compute the quasiparticle energies of bulk 2H-TMDCs by using the \textsc{yambo} code \cite{yambo}. The dielectric screening $\epsilon_{\vb{G},\vb{G}'}(\vb{q},\omega)$ is obtained considering the Plasmon-Pole approximation in a $\vb{k}$-grid of $12\times12\times4$,  with an energy cutoff of 60 Ry for the exchange potential and 30 Ry for the screened interaction $W_0$. Both the Green's functions and the screened Coulomb interaction are calculated including 600 bands. This number of bands is sufficient to provide accurate results thanks to the implementation of the Bruneval-Gonze terminators approach \cite{BGterminatorPRB2008}.

The electron-hole (e-h) interaction is included
in the calculation of the optical absorption spectra of the 2H-TMDCs by means of the
Bethe-Salpeter equation (BSE), which can be written  adopting the Tamm-Dancoff approximation as an effective eigenvalue problem
\begin{equation}
\sum_{v'c'\vb{k}'}H^{\text{BSE}}_{\substack{vc\vb{k}\\v'c'\vb{k}'}}A^{S}_{v'c'\vb{k}'} = \Omega_SA^{S}_{vc\vb{k}}\,,
\end{equation}
where $A^S_{vc\mathbf{k}}$ and $\Omega_S$ are the exciton eigenfunction and the eigenvalues for the $S$-th exciton, respectively. In this form, the BSE Hamiltonian is given by
\begin{equation}
H^{\text{BSE}}_{\substack{vc\vb{k}\\v'c'\vb{k}'}} = \left(\epsilon^{\text{QP}}_{c\mathbf{k}}-\epsilon^{\text{QP}}_{v\mathbf{k}}\right)\var_{c,c'}\var_{v,v'}\var_{\vb{k},\vb{k}'}+ \Xi_{vc\vb{k},v'c'\vb{k}'}\,,
\end{equation}
where the first term comprises a diagonal part that contains the
quasiparticle energy differences, and $\Xi  = \bar{v} - W$ is the e-h interaction kernel, composed of  a bare, repulsive
short-range exchange term, and an attractive static screened Coulomb
potential \cite{louie2000PRB}. 

To speed up the BSE calculations, we employ the double-grid method \cite{kammerlander2012} in which the kernel matrix elements are calculated on a coarse \textbf{k}-grid (the same used for the $GW$ calculations), and then interpolated on a finer \textbf{k}-grid sampling of 60 $\times$ 60 $\times$ 8, including eight valence bands and eight conduction bands. These values are sufficient to obtain converged results for the lowest peaks of the optical absorption. A full analysis of the convergence of all the relevant parameters related to $GW$ and BSE calculations is presented in the Supporting Information \cite{SI}.  

The optical absorption is calculated considering the imaginary
part of the dielectric function
\begin{equation}
    \epsilon_2(\omega) = \frac{4\pi^2}{VN_k}\sum_{S} \abs{\sum_{vc\vb{k}} A^S_{cv\vb{k}}\hat{\vb{e}}\cdot\mel{v\vb{k}}{\vb{r}}{c\vb{k}}}^2\var(\omega - \Omega_S)\,,\label{eq:epsilon2}
\end{equation}
where $V$ is the volume, $N_k$ is the number of points in the Brillouin zone sampling, $\hat{\vb{e}}$ is the light polarization vector, and $\mel{v\vb{k}}{\vb{r}}{c\vb{k}}$ are the single-particle dipole matrix elements.


\section{\label{sec:results}Results and Discussion}
\begin{figure}[h!]
    \centering
    \includegraphics[width=.6\textwidth]{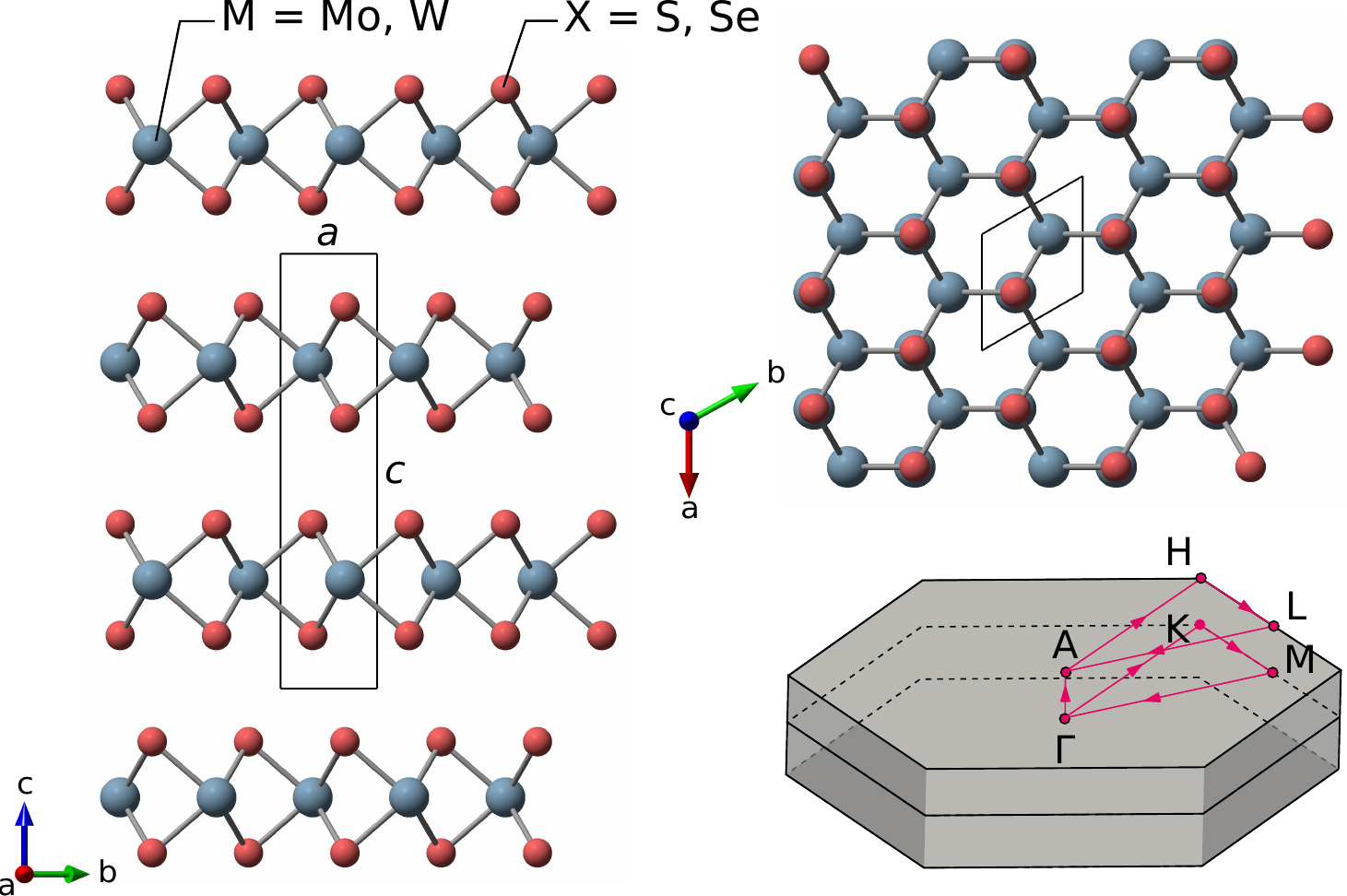}
    \caption{Top and side views of the crystal structure of bulk transition metal dichalcogenides \ce{MX2} in trigonal prismatic (2H) phase, where M (blue) is the transition metal Mo or W, and X (red) is the chalcogen atom S or Se. The first Brillouin zone is also represented with high-symmetry $\vb{k}$-points and paths.}
    \label{fig:1-structure}
\end{figure}
\begin{table}[t!]
    \centering
    \caption{Fully-relaxed lattice parameters for bulk 2H-TMDCs considering PBE-D2 van der Waals corrections and fully including spin-orbit coupling. The respective experimental values from Refs. \cite{wildervanck1964MoS2,Schutte1987WS2-WSe2,evans1971MoSe2} are shown for comparison.}
     \label{table1}
    \begin{tabular}{l cccc}
   \hline\hline
       & \ce{MoS2}&\ce{MoSe2}&\ce{WS2}&\ce{WSe2} \\\hline
       PBE-D2-SOC  &  &  & &  \\
       $a$ (\AA) & 3.19 & 3.32 &3.19 & 3.34  \\
       $c$ (\AA) & 12.42 & 13.03 &12.16 & 12.81  \\
       Experiment  &  &  & &  \\
       $a$ (\AA) & 3.16\textsuperscript{\textcolor{blue}{a}} & 3.29\textsuperscript{\textcolor{blue}{b}}  &3.15\textsuperscript{\textcolor{blue}{c}} & 3.28\textsuperscript{\textcolor{blue}{c}}  \\
       $c$ (\AA) & 12.29\textsuperscript{\textcolor{blue}{a}} & 12.93\textsuperscript{\textcolor{blue}{b}}  &12.32\textsuperscript{\textcolor{blue}{c}} & 12.96\textsuperscript{\textcolor{blue}{c}}  \\  \hline\hline
    \end{tabular}
    
    \begin{flushleft}
    \textsuperscript{\textcolor{blue}{a}} Ref.~\cite{wildervanck1964MoS2}; 
    \textsuperscript{\textcolor{blue}{b}} Ref.~\cite{evans1971MoSe2}; 
    \textsuperscript{\textcolor{blue}{c}} Ref.~\cite{Schutte1987WS2-WSe2}
    \end{flushleft}
\end{table}
 The computed fully-relaxed crystal structure is shown in Fig.~\ref{fig:1-structure}, and the lattice parameters of the studied bulk systems are summarized in Table \ref{table1}, which are in good agreement with experimental results \cite{wildervanck1964MoS2,Schutte1987WS2-WSe2,evans1971MoSe2}.

\begin{figure}[h!]
    \centering
    \includegraphics[width=\linewidth]{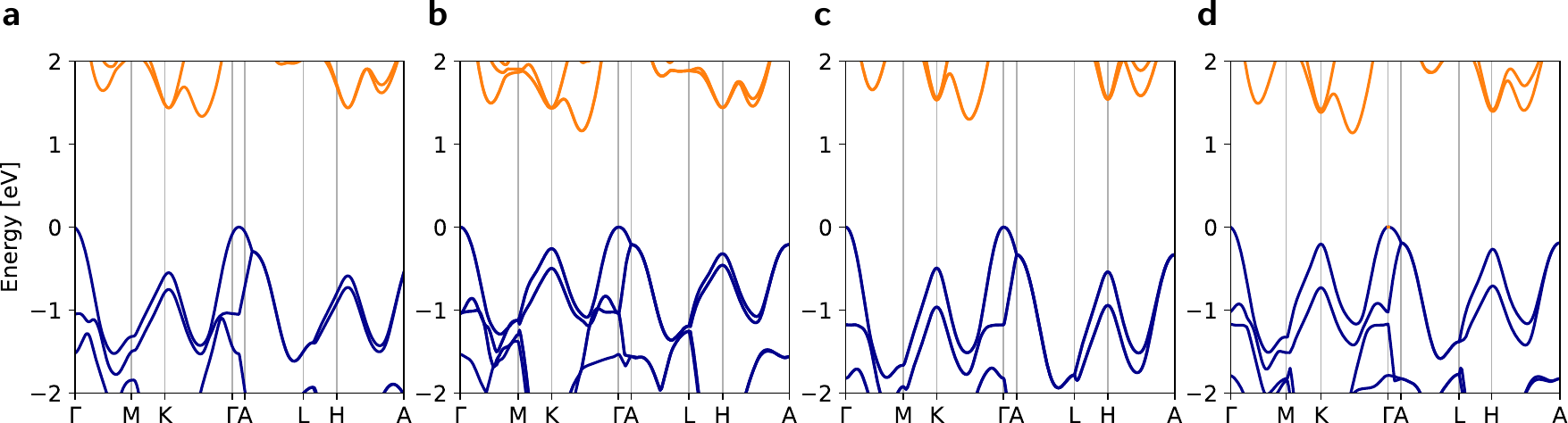}
    \caption{\Hightlight{Quasiparticle band structure of (a) \ce{MoS2}, (b) \ce{MoSe2}, (c) \ce{WS2} and (d) \ce{WSe2} within $G_0W_0$ approximation including spin-orbit coupling.}}
    \label{fig:gw-bands}
\end{figure}

Subsequently, \Hightlight{we compute the quasiparticle band structures and band gaps of the bulk TMDCs using the $G_0W_0$ approximation including spin-orbit coupling. The results are presented in Fig.~\ref{fig:gw-bands} and Table \ref{table2-gaps-eb}.}   \Hightlight{Our results, which include both direct and indirect quasiparticle band gap values, align well with the previously reported $G_0W_0$ findings \cite{Kim2021PRB}. Moreover, our calculations are in agreement with the band gaps and spin-splitting observed for the valence band, denoted as $\Delta_V$, as measured through ARPES according to Kim et al. \cite{kim2016determination}.} It can be seen that the band gaps of the TMDCs are similar for those systems with the same chalcogen species. 

\begin{table}[t!]
    \centering
    \caption{\Hightlight{Quasiparticle fundamental (indirect) and direct band gap (in eV) at $K$ point in the Brillouin zone, calculated within $G_0W_0$ approximation; spin-splitting for the valence band $\Delta_V$ (in meV);  ground-state and A excitonic peak energy positions (X$_{\text{gs}}$, X$_A$), and Boltzmann factor $\Delta_{\text{X}}$ for the fraction of the radiative e-h recombination  in the SLME-X approach for the bulk 2H-TMDCs (see text for further details). For comparison, we present the band gaps and spin-splitting $\Delta_V$ obtained by ARPES \cite{kim2016determination}.}}
     \label{table2-gaps-eb}
     \renewcommand{\arraystretch}{1.2}
    \begin{tabular}{l ccc ccc}
   \hline\hline
       &$E_{\text{g}}^{i}$& $E^{d}_{\text{g}}$& $\Delta_V$ & X$_{\text{gs}}$&X$_A$&$\Delta_{\text{X}}$ \\\hline
      \multirow{2}{*}{\ce{MoS2}}  & 1.34 & 1.99 & 0.20 & 1.846 & 1.877 & 0.031 \\
       &1.14\textsuperscript{\textcolor{blue}{a}} &1.82\textsuperscript{\textcolor{blue}{a}} & 0.16\textsuperscript{\textcolor{blue}{a}} & & & \\ 
       \multirow{2}{*}{\ce{MoSe2}}& 1.16 & 1.69 & 0.24 & 1.583 & 1.607 & 0.024 \\
       &1.25\textsuperscript{\textcolor{blue}{a}} &1.57\textsuperscript{\textcolor{blue}{a}} & 0.20\textsuperscript{\textcolor{blue}{a}} & & & \\ 
       \multirow{2}{*}{\ce{WS2}}  & 1.30 & 2.02 &0.47 & 1.915 & 1.956 & 0.041 \\
       &1.25\textsuperscript{\textcolor{blue}{a}} &1.82\textsuperscript{\textcolor{blue}{a}} & 0.44\textsuperscript{\textcolor{blue}{a}} & & & \\ 
       \multirow{2}{*}{\ce{WSe2}}  & 1.14 & 1.59 & 0.49& 1.511 & 1.561 & 0.050 \\
       &1.12\textsuperscript{\textcolor{blue}{a}} &1.62\textsuperscript{\textcolor{blue}{a}} & 0.48\textsuperscript{\textcolor{blue}{a}} & & & \\ 
        \hline\hline
    \end{tabular}
    \begin{flushleft}
    \textsuperscript{\textcolor{blue}{a}} Ref.~\cite{kim2016determination}
    \end{flushleft}
\end{table}

\begin{figure*}[t]
    \centering
    \includegraphics[width=\textwidth]{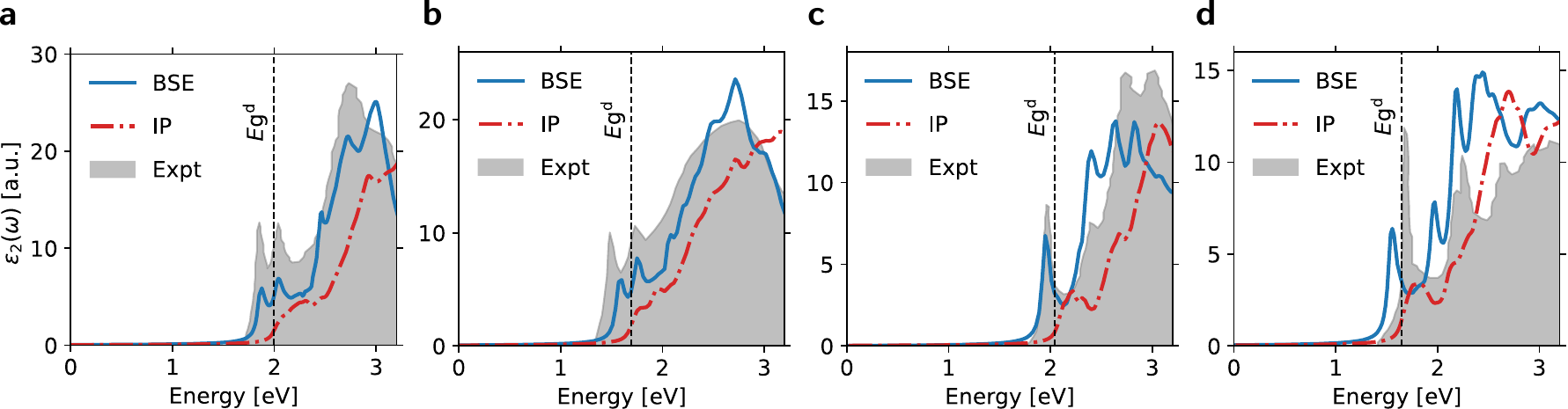}
    \caption{ Optical spectra calculated at the BSE (blue solid line) and independent particle (IP, red dot-dashed line) levels for the bulk (a) \ce{MoS2}, (b) \ce{MoSe2}, (a) \ce{WS2}, and (a) \ce{WSe2}. The Kohn-Sham eigenvalues include the quasiparticle corrections of the $G_0W_0$ calculations. The experimental data (gray shadow areas) are taken from Refs.~\cite{beal1976,beal1979}.}
    \label{fig:2-optical-spectra}
\end{figure*}

Turning to the optical properties, the computed imaginary part of the dielectric function at both the independent particle (IP) level and including e-h interaction are shown in Fig.~\ref{fig:2-optical-spectra}. For comparison, we also show the experimental optical absorption results (Refs.~\cite{beal1976} and \cite{beal1979}) for the respective TMDCs (shadow areas).

\Hightlight{We notice a good agreement between the BSE optical spectra and the experimental ones, highlighting the excitons' role in the optical absorption of bulk TMDCs. Overall, our findings effectively capture the essential characteristics of the experimental spectra, albeit with slightly diminished accuracy for W-based TMDCs primarily attributable to the increased splitting of electronic bands induced by higher spin-orbit coupling. The lack of prominent peaks in the IP spectrum strongly suggests that the sharp redshifted X peaks observed in BSE are associated with excitonic features. The absorption spectra calculated within IP exhibit noticeably weaker and less defined shoulder-like characteristics, indicating primarily band transitions.} The two lowest energy peaks correspond to the excitonic states labeled as A and B, which are associated with interband transitions at the $K$ point in the Brillouin zone, \textcolor{black}{coming} from the splitting of the valence band due to spin-orbit coupling \cite{li2014measurement,wang2018colloquium}. 

Our results indicate that the ground-state exciton (X$_{\text{gs}}$) for all the studied systems corresponds to a dark state, that rises from transitions between the valence band maximum and conduction band minimum at the $K$ point. The corresponding excitonic peak energy position of X$_{\text{gs}}$ and X$_A$ for the TMDCs are summarized in Table \ref{table2-gaps-eb}.

Now we investigate the photovoltaic efficiency of the bulk TMDCs. Yu and Zunger \cite{Yu-Zunger-PRL-SLME} developed the so-called spectroscopic limited maximum efficiency (SLME) model to compute the maximum efficiency of the absorber in a single-junction solar cell. In this formalism, the magnitude of the fundamental band gap, the values of the direct and indirect electronic gaps, the standard solar spectrum, the absorption coefficient of the material, and the thickness are taken into account to estimate the PCE of the device
\begin{equation}
    \eta_{\text{SLME}} = \frac{P_{\text{m}}}{P_{\text{in}}}\,,
\end{equation}
where $P_{\text{in}}$ is
the total incident power density with a value of $1000.37$ W/m$^2$, and $P_{\text{m}}$ is the maximum output power density which
can be obtained by maximizing the product of current density, $J$, and voltage, $V$,
\begin{equation}
    P = JV = \left[J_{\text{sc}} - J_0(e^{eV/k_{\text{B}}T}-1)\right]V\,,
\end{equation}
where $k_{\text{B}}$ is the Boltzmann constant, and $T$ is the device temperature
taken as $T = 298.15$ K throughout this work. 
Both the short circuit current density $J_{sc}$ and the
reverse saturation current density $J_0$ are calculated from
the photon absorptivity $a(\omega)$ of the material as follows
\begin{equation}
    a(\omega) = 1 - e^{-2\alpha(\omega)L}\,,
\end{equation}
where  $\alpha(\omega)$ is the absorption coefficient computed from first principles, and
$L$ is the thickness of the photovoltaic absorber material taken as 500 nm unless otherwise stated. In this regard, those current densities can be described including the AM1.5G solar spectrum $I_{\text{sun}}(\omega)$ and the black-body
spectrum $I_{\text{bb}}(\omega,T)$, respectively
\begin{eqnarray}
    J_{\text{sc}} &=&e\int_{E_\text{g}}^\infty a(\omega)I_{\text{sun}}(\omega)\,\dd\omega\,,\label{eq:jsc}\\
    J_{\text{0}} &=& \frac{J_0^{\text{r}}}{f_{\text{r}}}=\frac{e\pi}{f_{\text{r}}} \int_{E_\text{g}}^\infty a(\omega)I_{\text{bb}}(\omega,T)\,\dd\omega\,.\label{eq:j0}
\end{eqnarray}

$J_0$ corresponds to the total e-h recombination current at equilibrium in the dark, regarding both radiative and non-radiative recombinations. Here, $f_{\text{r}}$ is proposed as the fraction of the radiative recombination current, which in the original SLME framework is modeled using a Boltzmann factor
\begin{equation}
    f_{\text{r}} = e^{-\Delta_r/k_{\text{B}}T}\,,\label{eq:fr}
\end{equation}
where $\Delta_r\equiv E_\text{g}^{\text{da}}-E_\text{g}$ is the difference between the allowed direct band gap and the fundamental band gap, respectively.

Note that the efficiency within the Shockley–Queisser (SQ) detailed balance limit \cite{sq-limit} ($\eta_{\text{SQ}}$) can be obtained from the equations above by setting $f_r=1$ and considering the absorbance in Eq.~(\ref{eq:jsc}) and (\ref{eq:j0}) as a Heaviside step function that vanishes for $E<E_\text{g}$ and is equal to $1$ for $E\geq E_\text{g}$, where $E_\text{g}$ is the optical band gap of the photovoltaic absorber.

\begin{figure*}[t!]
    \centering
    \includegraphics[width=.85\textwidth]{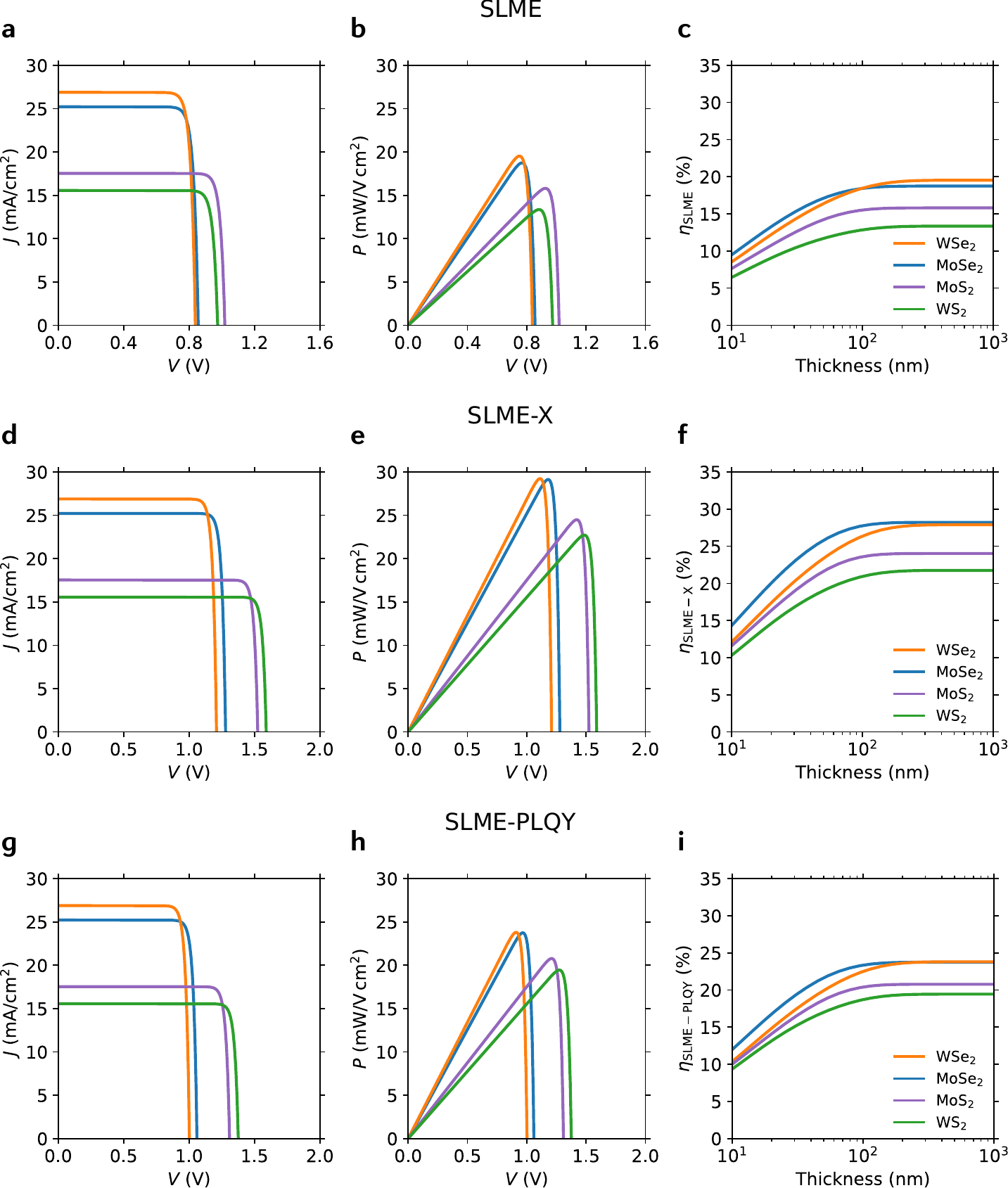}
    \caption{\Hightlight{Photovoltaic properties of bulk 2H-TMDCs calculated using the spectroscopic limited maximum efficiency (SLME) framework (a-c), modeling the fraction of radiative recombination losses ($f_r$) by considering the exciton eigenvalues from BSE (SLME-X) (d-f), and by extracting the fraction of the radiative losses directly from photoluminescence quantum yield experiments \cite{mohamed2017} (SLME-PLQY) (g-i): $J$-$V$, $P$-$V$ curves, and power conversion efficiency ($\eta$) as a function of the thickness of the absorber. The results have been computed for a device temperature of $T=298.15$ K.}}
    \label{fig:3-slme}
\end{figure*}
\begin{table*}[th!]
    \centering
    \caption{\Hightlight{Photovoltaic properties of bulk TMDCs obtained within the Shockley-Queisser approach and the spectroscopic limited maximum efficiency (SLME) standard framework, by modeling the fraction of radiative recombination losses ($f_r$) by considering the exciton eigenvalues from BSE (SLME-X), and by extracting the fraction of the radiative losses directly from photoluminescence quantum yield experiments \cite{mohamed2017} (SLME-PLQY): short-circuit current densities ($J_\text{sc}$), open-circuit voltage ($V_\text{oc}$), fill factor (FF), and power-conversion efficiency ($\eta$). }}
     \label{table3-efficiencies}
     \renewcommand{\arraystretch}{1.2}
    \resizebox{\textwidth}{!}{%
    \begin{tabular}{l cccc ccccc}
   \hline\hline
       & \multicolumn{4}{c}{Shockley-Queisser} & & \multicolumn{4}{c}{SLME}\\\cline{2-5}\cline{7-10}
       &$J_{\text{sc}}$ $\left(\right.$mA/cm$^2$$\left.\right)$& $V_{\text{oc}}$ (V) & FF (\%) & $\eta_{\text{SQ}}$ (\%) & &$J_{\text{sc}}$ $\left(\right.$mA/cm$^2$$\left.\right)$ &$V_{\text{oc}}$ (V) & FF (\%)&$\eta_{\text{SLME}}$ (\%) \\\hline
       \ce{MoS2}  & 17.5 & 1.55 & 92.2 & 25.0 & & 17.5 & 1.02 & 88.6 & 15.8 \\
       \ce{MoSe2} &25.2 & 1.30 & 90.7 & 29.7 & & 25.2 & 0.86 & 86.5 & 18.7  \\
       \ce{WS2}   & 15.6 & 1.63 & 91.8 & 23.3 & & 15.6 & 0.98 & 87.4 & 13.4  \\
       \ce{WSe2}  & 26.9 & 1.26 & 90.1 & 30.5 & & 26.9 & 0.84 & 86.4 & 19.5  \\[.5cm]
       & \multicolumn{4}{c}{SLME-X} & & \multicolumn{4}{c}{SLME-PLQY}\\\cline{2-5}\cline{7-10}
       &$J_{\text{sc}}$ $\left(\right.$mA/cm$^2$$\left.\right)$& $V_{\text{oc}}$ (V) & FF (\%) & $\eta_{\text{SLME-X}}$ (\%) & &$J_{\text{sc}}$ $\left(\right.$mA/cm$^2$$\left.\right)$ &$V_{\text{oc}}$ (V)& FF (\%)&$\eta_{\text{SLME-PLQY}}$ (\%) \\\hline
       \ce{MoS2}  & 17.5 & 1.52 & 92.0 & 24.5 & & 17.5 & 1.31 & 90.5 & 20.8 \\
       \ce{MoSe2} &25.2 & 1.28 & 90.3 & 29.1 & & 25.2 & 1.06 & 88.8 & 23.7  \\
       \ce{WS2}   & 15.6 & 1.59 & 91.6 & 22.7 & & 15.6 & 1.37 & 91.1 & 19.4  \\
       \ce{WSe2}  & 26.9 & 1.21 & 89.8 & 29.2 & & 26.9 & 1.01 & 88.1 & 23.8  \\
        \hline\hline 
    \end{tabular}}
\end{table*}

It is important to notice that for indirect band gap materials with large $\Delta_r$, as in the TMDCs studied here, the original SLME approach can drastically underestimate the predicted PCE, given that the Boltzmann factor rapidly decreases the fraction of radiative recombination current density $f_r$. Consequently, the saturation current $J_0$ is overestimated, yielding lower values for the open-circuit voltages. This effect was studied in detail by Bercx \emph{et al.} \cite{bercx2016first} for silicon, which possesses a $\Delta_r=2.23$ eV. This leads to a PCE of zero regardless of the temperature and thickness of the material.

First, we apply the standard SLME formalism to compute the PCEs considering $\Delta_r$ as the difference between $X_{A}$ and $E_\text{g}^{i}$. Employing this approach, for all the analyzed systems,
 the fraction of radiative recombination current, $f_r$, is much smaller than 10$^{-8}$, leading to 
PCEs \Hightlight{varying between 13\% and 19\% for the studied TMDCs, significantly smaller than those obtained through Shockley-Queisser (Table \ref{table3-efficiencies})}. This result suggests that, for these materials, employing the original SLME approach implies that almost all e-h recombination would be non-radiative, even though $J_0$ is modeled assuming a black-body spectrum, which relies on a radiative principle. This issue motivated us to explore alternative models for describing the recombination factor.

Delving into this possibility, and taking into account that the excitonic effects in bulk TMDCs play a key role in their optical responses,  we also compute the PCE using a slightly modified SLME approach that has been proposed by Ozório and co-workers \cite{ozorio2022theoretical}. In their model, the fraction of the radiative recombination current, $f_{\text{r}}$, is calculated  by replacing the $\Delta_r$ in Eq. (9) with $\Delta_\text{X}$, which is defined as
\begin{equation}
    \Delta_{\text{X}} = \text{X}_A - \text{X}_{\text{gs}}.
\end{equation} 
We refer to this approach as exciton-based SLME or SLME-X. Within it, the fraction of the radiative recombination current, $f_r$, calculated taking into account the $\Delta_{\text{X}}$ values presented in Table \ref{table2-gaps-eb}. As a result, the obtained $f_r$ values vary between 1.5$\times$10$^{-2}$ and 3.9$\times$10$^{-2}$. \Hightlight{It is important to point out that the presence of an indirect gap could lead to lower energy dark excitons that lead to non-radiative decay. Thus, the most effective method for obtaining the exciton ground state involves computing the $q$-dependent BSE spectra \cite{moujaes2023}. In that case, we expect that $f_r$ would have even lower values, although still surpassing those obtained within the standard SLME.}


In Fig.~\ref{fig:3-slme}a-b, we show the current and power density curves as a function of the voltage, obtained with the SLME-X approach. The PCE as a function of thickness is depicted in Fig.~\ref{fig:3-slme}c. All the main solar cell parameters obtained from these curves are summarized in Table \ref{table3-efficiencies}.

We can verify that all the investigated photovoltaic properties are similar for those TMDCs that are composed of the same chalcogen species. The transition metal diselenides should have higher photovoltaic efficiencies than disulfides due to their lower optical band gaps. \textcolor{black}{In addition, our results presented in Fig.~\ref{fig:3-slme}\textcolor{blue}{(c)} indicate that $\eta_{\text{SLME-X}}$ saturates for a thickness around $10^2$ nm. }

\textcolor{black}{Alternatively,} non-radiative recombination losses can be experimentally quantiﬁed by the photoluminescence quantum yield (PLQY) \cite{Dreessen23}. We can also estimate the maximum PCE by assuming that the PLQY is, in principle, equal to the fraction of radiative recombination ($f_{r}$) \cite{jariwala2017van}. We denote this other modified SLME approach as SLME-PLQY. Experimental studies have reported that for single-layer TMDCs, the PLQY is close to unity \cite{amani2015near,amani2016recombination,mohamed2017}. In contrast, it has been demonstrated that the PLQY for their bulk counterpart is reduced by a factor of 10$^{4}$ \cite{mak2010atomically}. 

We adopt the PLQY values reported by Mohamed and co-workers \cite{mohamed2017} as the experimental-based $f_{r}$ for the TMDCs. These $f_r$ values are 6.8$\times$10$^{-5}$, 7.9$\times$10$^{-5}$, 4.9$\times$10$^{-5}$, and 4.5$\times$10$^{-5}$ for the bulk \ce{MoS2}, \ce{MoSe2}, \ce{WS2} and \ce{WSe2}, respectively. 

Considering the SLME-PLQY approach, we also calculate the current density-voltage, power density-voltage, and PCE-thickness curves of the studied TMDCs, which are shown in Fig.~\ref{fig:3-slme}d-f. For comparison with the SLME-X results, we describe the solar cell parameters obtained within SLME-PLQY for the TMDCs in Table \ref{table3-efficiencies}. 
\textcolor{black}{Given that the short-circuit current densities, $J_{sc}$, do not depend on the fraction of the radiative recombination current, their values are identical for both approaches. On the other hand, the open-circuit voltages and fill factors are smaller in the SLME-PLQY approach in comparison to the SLME-X one, leading to smaller values of PCE in the latter case.}

\begin{figure}[t!]
    \centering
    \includegraphics[width=.5\textwidth]{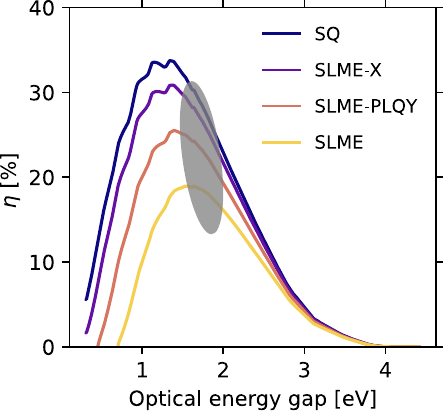}
    \caption{Power conversion efficiency with respect to the optical energy gap in Shockley-Queisser and SLME-based approaches. The plotted curves, derived from the average computed fraction of e-h radiative recombination ($f_r$) for the studied TMDCs, illustrate the impact of nonradiative recombination losses in the PCE. Alongside the standard SLME, we incorporate considerations from exciton eigenvalues obtained through BSE (SLME-X) and direct estimations from photoluminescence quantum yield analyses experiments (SLME-PLQY). The shaded area denotes the range of optical band gaps associated with the analyzed TMDCs.}
    \label{fig:4-pce-optical-eg}
\end{figure}

All the obtained results of the photovoltaic efficiencies within the SQ and SLME-based approaches are summarized in Fig.~\ref{fig:4-pce-optical-eg}, where we present the maximum PCE as a function of the optical band gap. For the analyzed TMDCs, the average of the fraction of e-h radiative recombination is $f_r = 3\times10^{-8}$ for the standard SLME, $f_r = 6\times10^{-5}$ for SLME-PLQY and $f_r = 3\times10^{-2}$ for SLME-X. \Hightlight{In addition, we have estimated the error associated with those average maximum PCEs, by introducing in the $f_r$ the error threshold of the $GW$ calculations (see Fig.~S5 in the Supporting Information). As a result, regarding the SLME approach, for systems with optical gaps close to 1.9 eV (\ce{MoS2}, \ce{WS2}), the estimated error is ${\sim}$1\%, whereas for those with optical gaps around 1.6 eV (\ce{MoSe2}, \ce{WSe2}), the error can be approximately 2.4\%. Considering this uncertainty estimate, we expect comparable errors for the remaining formalisms.} 

Based on these fractions of non-radiative e-h recombination, we compute the efficiencies as a function of the optical energy gap. The shaded area in Fig.~\ref{fig:4-pce-optical-eg} represents the range of optical band gaps corresponding to the TMDCs studied. We can verify that the SQ formalism, which considers only radiative recombination processes ($f_r = 1$), overestimates the photovoltaic efficiency of the bulk TMDCs, with photovoltaic efficiencies ranging from ${\sim}23$\% to ${\sim}29$\% (Table S1 in the Supporting Information \cite{SI}). On the other hand, all the SLME-based approaches include a non-vanishing fraction of non-radiative e-h recombination ($f_r < 1$), decreasing the predicted photovoltaic efficiencies of the absorbers.

Our results suggest that even the highest quality TMDC-based thin film would yield maximum efficiencies between ${\sim}23-29$\%, if we take into account only excitonic effects as a fundamental factor to describe the recombination processes. Nevertheless, a more realistic upper limit of the photovoltaic efficiency for the investigated TMDCs can be predicted by considering the fraction of the radiative recombination directly from PLQY experiments, with the TMDC thin-films reaching maximum PCEs between ${\sim}19-21$\%. \textcolor{black}{Based on it, it is noticeable that PLQY-based efficiencies, reflecting the performance of impurity-containing samples, are inferior to those derived from considering only excitonic effects. This implies that there is still room for improving the TMDC samples to attain greater PCEs. }

The obtained efficiencies can be compared with previous theoretical studies. For instance,  Ozdemir and Baroni \cite{ozdemir2020thickness} investigated the photovoltaic efficiency of bulk TMDCs based on ab initio many-body perturbation theory calculations, including the e-h interaction. In their work, the absorptivity is calculated as $A=1-T-R$, where $T$ is the transmitivity and $R$ is the reflectivity, and those properties are computed from Fresnel’s equations. For TMDCs with a thickness of $10^3$ nm, the same ones analyzed in our work, they reported values of PCE in a range of $\sim$ 21\%  to $\sim$ 26\%, being the selenides the most efficient systems, consistent with our findings, especially those that account for nonradiative recombination through PLQY experiments.

Furthermore, Chouldhary et al. \cite{choudhary2019accelerated} have reported theoretical results of SLME based on DFT without taking e-h interactions into account. The $\eta_{\text{SLME}}$ computed for a thickness of 500 nm were $18.6\%$, $27.6\%$, $15.2\%$, and $21.5\%$ for bulk \ce{MoS2}, \ce{MoSe2}, \ce{WS2}, and \ce{WSe2}, respectively. Although in that work the authors have used band gaps calculated at the DFT level with meta-GGA functionals, their SLME efficiencies lie between our predicted values modeled by the SLME-X and SLME-PLQY approaches.

Finally, despite the large range of efficiencies, we verify that the selenides present higher efficiencies, with \ce{MoSe2} and \ce{WSe2} photovoltaic efficiencies being almost the same. Our results highlight the importance of the fraction of radiative e-h recombination factor ($f_r$) to determine the maximum PCEs of bulk TMDCs, which indeed can be extrapolated to other indirect band gap absorbers. In particular, we present a range of values for this relevant solar cell parameter and their corresponding efficiencies, based on several approaches that can represent reliable upper limits for the photovoltaic efficiencies of single-junction solar devices produced with TMDC thin films.


\section{\label{sec:conclusions}Conclusions}
We carried out first-principles calculations based on DFT combined with many-body perturbation theory within the $G_0W_0$-BSE framework to study the optical response and the photovoltaic properties of bulk 2H-\ce{MX2} (M$\,=\,$Mo, W, and X$\,=\,$S, Se), using for the latter the spectroscopic limited maximum efficiency (SLME)
 formalism.  Our results indicated that an accurate description of the optical spectrum within the $GW$-BSE framework requires mapping out the Brillouin zone in very fine \emph{k}-grid samples. The agreement obtained when comparing the BSE optical spectra and the
experimental ones highlights the excitons’ roles in the optical absorption of bulk TMDCs.  The calculated power conversion efficiencies (PCEs) of the analyzed TMDCs were overestimated with the Shockley-Queisser formalism, as expected since in this approach one includes only radiative e-h recombination losses. On the other hand, the standard SLME approach tends to underestimate the PCE of these van der Waals materials, due to a quite small estimate of the fraction of radiative recombination. In other words, for the standard SLME, almost all the recombination processes should correspond to nonradiative losses. These findings have motivated us to explore alternative methods for modeling PCEs that account for radiative and nonradiative recombination processes in more realistic ways. Thus, we investigate the modeling of the fraction of e-h recombination processes by considering the energy range from exciton eigenvalues (SLME-X), and also by employing experimental results obtained for photoluminescence quantum yield (SLME-PLQY). 
As a result, we predict that when we adopt the exciton energy ranges to compute the PCE within the SLME-X approach, the maximum PCEs vary between ${\sim}23-29$\%, whereas taking the experimental fraction of radiative e-h recombination from PLQY experiments within the SLME-PLQY approach, the bulk TMDC  maximum PCEs reached values between ${\sim}19-24$\%.
Regardless of the approach used to determine the PCE,
our results indicate that the selenide-based TMDCs exhibit superior photovoltaic efficiency in comparison to the sulfide-based ones.
In conclusion, we believe that our comprehensive analysis of the effects of the fraction of e-h radiative recombination in the TMDC efficiencies can be used as accurate predictions of upper limits for the photovoltaic efficiencies of solar devices based on few-layer or multi-layer TMDCs, paving the way for the development of efficient nanoscopically thin solar cells.

\section*{\label{sec:acknowledgements}Acknowledgements}
E.M.Jr and A.R.R. acknowledge the financial support from the Brazilian agency FAPESP, Grant No. 20/13172-8 and 2017/02317-2. This research was supported by resources supplied by CENAPAD-SP and the Center for Scientific Computing (NCC/GridUNESP) of the UNESP.

\bibstyle{achemso}
\bibliography{paper}

\providecommand{\latin}[1]{#1}
\makeatletter
\providecommand{\doi}
  {\begingroup\let\do\@makeother\dospecials
  \catcode`\{=1 \catcode`\}=2 \doi@aux}
\providecommand{\doi@aux}[1]{\endgroup\texttt{#1}}
\makeatother
\providecommand*\mcitethebibliography{\thebibliography}
\csname @ifundefined\endcsname{endmcitethebibliography}  {\let\endmcitethebibliography\endthebibliography}{}
\begin{mcitethebibliography}{63}
\providecommand*\natexlab[1]{#1}
\providecommand*\mciteSetBstSublistMode[1]{}
\providecommand*\mciteSetBstMaxWidthForm[2]{}
\providecommand*\mciteBstWouldAddEndPuncttrue
  {\def\EndOfBibitem{\unskip.}}
\providecommand*\mciteBstWouldAddEndPunctfalse
  {\let\EndOfBibitem\relax}
\providecommand*\mciteSetBstMidEndSepPunct[3]{}
\providecommand*\mciteSetBstSublistLabelBeginEnd[3]{}
\providecommand*\EndOfBibitem{}
\mciteSetBstSublistMode{f}
\mciteSetBstMaxWidthForm{subitem}{(\alph{mcitesubitemcount})}
\mciteSetBstSublistLabelBeginEnd
  {\mcitemaxwidthsubitemform\space}
  {\relax}
  {\relax}

\bibitem[Han(2020)]{vdW-Science}
Han,~X. Ductile van der Waals materials. \emph{Science} \textbf{2020}, \emph{369}, 509--509\relax
\mciteBstWouldAddEndPuncttrue
\mciteSetBstMidEndSepPunct{\mcitedefaultmidpunct}
{\mcitedefaultendpunct}{\mcitedefaultseppunct}\relax
\EndOfBibitem
\bibitem[Novoselov \latin{et~al.}(2004)Novoselov, Geim, Morozov, Jiang, Zhang, Dubonos, Grigorieva, and Firsov]{Novoselov2004}
Novoselov,~K.~S.; Geim,~A.~K.; Morozov,~S.~V.; Jiang,~D.; Zhang,~Y.; Dubonos,~S.~V.; Grigorieva,~I.~V.; Firsov,~A.~A. Electric Field Effect in Atomically Thin Carbon Films. \emph{Science} \textbf{2004}, \emph{306}, 666--669\relax
\mciteBstWouldAddEndPuncttrue
\mciteSetBstMidEndSepPunct{\mcitedefaultmidpunct}
{\mcitedefaultendpunct}{\mcitedefaultseppunct}\relax
\EndOfBibitem
\bibitem[Khan \latin{et~al.}(2020)Khan, Tareen, Aslam, Wang, Zhang, Mahmood, Ouyang, Zhang, and Guo]{khan2020}
Khan,~K.; Tareen,~A.~K.; Aslam,~M.; Wang,~R.; Zhang,~Y.; Mahmood,~A.; Ouyang,~Z.; Zhang,~H.; Guo,~Z. Recent developments in emerging two-dimensional materials and their applications. \emph{J. Mater. Chem. C} \textbf{2020}, \emph{8}, 387--440\relax
\mciteBstWouldAddEndPuncttrue
\mciteSetBstMidEndSepPunct{\mcitedefaultmidpunct}
{\mcitedefaultendpunct}{\mcitedefaultseppunct}\relax
\EndOfBibitem
\bibitem[Ma \latin{et~al.}(2021)Ma, Ren, Xu, and Ou]{ma2021}
Ma,~Q.; Ren,~G.; Xu,~K.; Ou,~J.~Z. Tunable optical properties of 2D materials and their applications. \emph{Adv. Opt. Mater.} \textbf{2021}, \emph{9}, 2001313\relax
\mciteBstWouldAddEndPuncttrue
\mciteSetBstMidEndSepPunct{\mcitedefaultmidpunct}
{\mcitedefaultendpunct}{\mcitedefaultseppunct}\relax
\EndOfBibitem
\bibitem[Villegas and Rocha(2022)Villegas, and Rocha]{villegas2022}
Villegas,~C. E.~P.; Rocha,~A.~R. Near-Infrared Optical Response and Carrier Dynamics for High Photoconversion in Tellurene. \emph{J. Phys. Chem. C} \textbf{2022}, \emph{126}, 6129--6134\relax
\mciteBstWouldAddEndPuncttrue
\mciteSetBstMidEndSepPunct{\mcitedefaultmidpunct}
{\mcitedefaultendpunct}{\mcitedefaultseppunct}\relax
\EndOfBibitem
\bibitem[Chhowalla \latin{et~al.}(2013)Chhowalla, Shin, Eda, Li, Loh, and Zhang]{chhowalla2013NatChem}
Chhowalla,~M.; Shin,~H.~S.; Eda,~G.; Li,~L.-J.; Loh,~K.~P.; Zhang,~H. The chemistry of two-dimensional layered transition metal dichalcogenide nanosheets. \emph{Nat. Chem.} \textbf{2013}, \emph{5}, 263--275\relax
\mciteBstWouldAddEndPuncttrue
\mciteSetBstMidEndSepPunct{\mcitedefaultmidpunct}
{\mcitedefaultendpunct}{\mcitedefaultseppunct}\relax
\EndOfBibitem
\bibitem[Manzeli \latin{et~al.}(2017)Manzeli, Ovchinnikov, Pasquier, Yazyev, and Kis]{Manzeli2017NatRevMat}
Manzeli,~S.; Ovchinnikov,~D.; Pasquier,~D.; Yazyev,~O.~V.; Kis,~A. {2D transition metal dichalcogenides}. \emph{Nat. Rev. Mater.} \textbf{2017}, \emph{2}, 17033--15\relax
\mciteBstWouldAddEndPuncttrue
\mciteSetBstMidEndSepPunct{\mcitedefaultmidpunct}
{\mcitedefaultendpunct}{\mcitedefaultseppunct}\relax
\EndOfBibitem
\bibitem[Perez \latin{et~al.}(2020)Perez, Amorim, Villegas, and Rocha]{perez2020}
Perez,~A.; Amorim,~R.~G.; Villegas,~C. E.~P.; Rocha,~A.~R. Nanogap-based all-electronic DNA sequencing devices using MoS 2 monolayers. \emph{Phys. Chem. Chem. Phys.} \textbf{2020}, \emph{22}, 27053--27059\relax
\mciteBstWouldAddEndPuncttrue
\mciteSetBstMidEndSepPunct{\mcitedefaultmidpunct}
{\mcitedefaultendpunct}{\mcitedefaultseppunct}\relax
\EndOfBibitem
\bibitem[Villegas and Rocha(2015)Villegas, and Rocha]{villegas2015}
Villegas,~C. E.~P.; Rocha,~A.~R. Elucidating the optical properties of novel heterolayered materials based on MoTe2--InN for photovoltaic applications. \emph{J. Phys. Chem. C} \textbf{2015}, \emph{119}, 11886--11895\relax
\mciteBstWouldAddEndPuncttrue
\mciteSetBstMidEndSepPunct{\mcitedefaultmidpunct}
{\mcitedefaultendpunct}{\mcitedefaultseppunct}\relax
\EndOfBibitem
\bibitem[Britnell \latin{et~al.}(2013)Britnell, Ribeiro, Eckmann, Jalil, Belle, Mishchenko, Kim, Gorbachev, Georgiou, Morozov, Grigorenko, Geim, Casiraghi, Neto, and Novoselov]{britnell-science}
Britnell,~L.; Ribeiro,~R.~M.; Eckmann,~A.; Jalil,~R.; Belle,~B.~D.; Mishchenko,~A.; Kim,~Y.-J.; Gorbachev,~R.~V.; Georgiou,~T.; Morozov,~S.~V.; Grigorenko,~A.~N.; Geim,~A.~K.; Casiraghi,~C.; Neto,~A. H.~C.; Novoselov,~K.~S. Strong Light-Matter Interactions in Heterostructures of Atomically Thin Films. \emph{Science} \textbf{2013}, \emph{340}, 1311--1314\relax
\mciteBstWouldAddEndPuncttrue
\mciteSetBstMidEndSepPunct{\mcitedefaultmidpunct}
{\mcitedefaultendpunct}{\mcitedefaultseppunct}\relax
\EndOfBibitem
\bibitem[Furchi \latin{et~al.}(2018)Furchi, H{\"o}ller, Dobusch, Polyushkin, Schuler, and Mueller]{furchi2018device}
Furchi,~M.~M.; H{\"o}ller,~F.; Dobusch,~L.; Polyushkin,~D.~K.; Schuler,~S.; Mueller,~T. Device physics of van der Waals heterojunction solar cells. \emph{npj 2D Mater. Appl.} \textbf{2018}, \emph{2}, 3\relax
\mciteBstWouldAddEndPuncttrue
\mciteSetBstMidEndSepPunct{\mcitedefaultmidpunct}
{\mcitedefaultendpunct}{\mcitedefaultseppunct}\relax
\EndOfBibitem
\bibitem[Baugher \latin{et~al.}(2014)Baugher, Churchill, Yang, and Jarillo-Herrero]{baugher2014optoelectronic}
Baugher,~B.~W.; Churchill,~H.~O.; Yang,~Y.; Jarillo-Herrero,~P. Optoelectronic devices based on electrically tunable p--n diodes in a monolayer dichalcogenide. \emph{Nat. Nanotechnol.} \textbf{2014}, \emph{9}, 262--267\relax
\mciteBstWouldAddEndPuncttrue
\mciteSetBstMidEndSepPunct{\mcitedefaultmidpunct}
{\mcitedefaultendpunct}{\mcitedefaultseppunct}\relax
\EndOfBibitem
\bibitem[Kim and Choi(2021)Kim, and Choi]{Kim2021PRB}
Kim,~H.~G.; Choi,~H.~J. {Thickness dependence of work function, ionization energy, and electron affinity of Mo and W dichalcogenides from DFT and GW calculations}. \emph{Phys. Rev. B} \textbf{2021}, \emph{103}, 1--9\relax
\mciteBstWouldAddEndPuncttrue
\mciteSetBstMidEndSepPunct{\mcitedefaultmidpunct}
{\mcitedefaultendpunct}{\mcitedefaultseppunct}\relax
\EndOfBibitem
\bibitem[Aftab \latin{et~al.}(2023)Aftab, Iqbal, Hussain, Hegazy, and Saeed]{Aftab23}
Aftab,~S.; Iqbal,~M.~Z.; Hussain,~S.; Hegazy,~H.~H.; Saeed,~M.~A. Transition metal dichalcogenides solar cells and integration with perovskites. \emph{Nano Energy} \textbf{2023}, \emph{108}, 108249\relax
\mciteBstWouldAddEndPuncttrue
\mciteSetBstMidEndSepPunct{\mcitedefaultmidpunct}
{\mcitedefaultendpunct}{\mcitedefaultseppunct}\relax
\EndOfBibitem
\bibitem[Nassiri~Nazif \latin{et~al.}(2021)Nassiri~Nazif, Kumar, Hong, Lee, Islam, McClellan, Karni, van~de Groep, Heinz, Pop, Brongersma, and Saraswat]{nassiri2021high}
Nassiri~Nazif,~K.; Kumar,~A.; Hong,~J.; Lee,~N.; Islam,~R.; McClellan,~C.~J.; Karni,~O.; van~de Groep,~J.; Heinz,~T.~F.; Pop,~E.; Brongersma,~M.~L.; Saraswat,~K.~C. High-performance p--n junction transition metal dichalcogenide photovoltaic cells enabled by MoO x doping and passivation. \emph{Nano Lett.} \textbf{2021}, \emph{21}, 3443--3450\relax
\mciteBstWouldAddEndPuncttrue
\mciteSetBstMidEndSepPunct{\mcitedefaultmidpunct}
{\mcitedefaultendpunct}{\mcitedefaultseppunct}\relax
\EndOfBibitem
\bibitem[He \latin{et~al.}(2022)He, Iwamoto, Kaneko, and Kato]{He2022}
He,~X.; Iwamoto,~Y.; Kaneko,~T.; Kato,~T. Fabrication of near-invisible solar cell with monolayer WS2. \emph{Sci. Rep.} \textbf{2022}, \emph{12}, 11315\relax
\mciteBstWouldAddEndPuncttrue
\mciteSetBstMidEndSepPunct{\mcitedefaultmidpunct}
{\mcitedefaultendpunct}{\mcitedefaultseppunct}\relax
\EndOfBibitem
\bibitem[Bernardi \latin{et~al.}(2013)Bernardi, Palummo, and Grossman]{Bernardi-NanoLett2013}
Bernardi,~M.; Palummo,~M.; Grossman,~J.~C. Extraordinary Sunlight Absorption and One Nanometer Thick Photovoltaics Using Two-Dimensional Monolayer Materials. \emph{Nano Lett.} \textbf{2013}, \emph{13}, 3664--3670\relax
\mciteBstWouldAddEndPuncttrue
\mciteSetBstMidEndSepPunct{\mcitedefaultmidpunct}
{\mcitedefaultendpunct}{\mcitedefaultseppunct}\relax
\EndOfBibitem
\bibitem[Britnell \latin{et~al.}(2013)Britnell, Ribeiro, Eckmann, Jalil, Belle, Mishchenko, Kim, Gorbachev, Georgiou, Morozov, Grigorenko, Geim, Casiraghi, Neto, and Novoselov]{britnell2013Science}
Britnell,~L.; Ribeiro,~R.~M.; Eckmann,~A.; Jalil,~R.; Belle,~B.~D.; Mishchenko,~A.; Kim,~Y.-J.; Gorbachev,~R.~V.; Georgiou,~T.; Morozov,~S.~V.; Grigorenko,~A.~N.; Geim,~A.~K.; Casiraghi,~C.; Neto,~A. H.~C.; Novoselov,~K.~S. Strong light-matter interactions in heterostructures of atomically thin films. \emph{Science} \textbf{2013}, \emph{340}, 1311--1314\relax
\mciteBstWouldAddEndPuncttrue
\mciteSetBstMidEndSepPunct{\mcitedefaultmidpunct}
{\mcitedefaultendpunct}{\mcitedefaultseppunct}\relax
\EndOfBibitem
\bibitem[Li \latin{et~al.}(2015)Li, Lee, Qu, Liu, Ryu, Seabaugh, and Yoo]{li2015ultimate}
Li,~H.-M.; Lee,~D.; Qu,~D.; Liu,~X.; Ryu,~J.; Seabaugh,~A.; Yoo,~W.~J. Ultimate thin vertical p--n junction composed of two-dimensional layered molybdenum disulfide. \emph{Nat. Comm.} \textbf{2015}, \emph{6}, 6564\relax
\mciteBstWouldAddEndPuncttrue
\mciteSetBstMidEndSepPunct{\mcitedefaultmidpunct}
{\mcitedefaultendpunct}{\mcitedefaultseppunct}\relax
\EndOfBibitem
\bibitem[Cheng \latin{et~al.}(2014)Cheng, Li, Zhou, Wang, Yin, Jiang, Liu, Chen, Huang, and Duan]{Cheng14}
Cheng,~R.; Li,~D.; Zhou,~H.; Wang,~C.; Yin,~A.; Jiang,~S.; Liu,~Y.; Chen,~Y.; Huang,~Y.; Duan,~X. Electroluminescence and Photocurrent Generation from Atomically Sharp WSe2/MoS2 Heterojunction p–n Diodes. \emph{Nano Lett.} \textbf{2014}, \emph{14}, 5590--5597\relax
\mciteBstWouldAddEndPuncttrue
\mciteSetBstMidEndSepPunct{\mcitedefaultmidpunct}
{\mcitedefaultendpunct}{\mcitedefaultseppunct}\relax
\EndOfBibitem
\bibitem[Mao \latin{et~al.}(2018)Mao, Zou, Li, Song, and He]{mao2018magnetron}
Mao,~X.; Zou,~J.; Li,~H.; Song,~Z.; He,~S. Magnetron sputtering fabrication and photoelectric properties of WSe2 film solar cell device. \emph{Appl. Surf. Sci.} \textbf{2018}, \emph{444}, 126--132\relax
\mciteBstWouldAddEndPuncttrue
\mciteSetBstMidEndSepPunct{\mcitedefaultmidpunct}
{\mcitedefaultendpunct}{\mcitedefaultseppunct}\relax
\EndOfBibitem
\bibitem[Went \latin{et~al.}(2019)Went, Wong, Jahelka, Kelzenberg, Biswas, Hunt, Carbone, and Atwater]{went2019new}
Went,~C.~M.; Wong,~J.; Jahelka,~P.~R.; Kelzenberg,~M.; Biswas,~S.; Hunt,~M.~S.; Carbone,~A.; Atwater,~H.~A. A new metal transfer process for van der Waals contacts to vertical Schottky-junction transition metal dichalcogenide photovoltaics. \emph{Sci. Adv.} \textbf{2019}, \emph{5}, eaax6061\relax
\mciteBstWouldAddEndPuncttrue
\mciteSetBstMidEndSepPunct{\mcitedefaultmidpunct}
{\mcitedefaultendpunct}{\mcitedefaultseppunct}\relax
\EndOfBibitem
\bibitem[McVay \latin{et~al.}(2020)McVay, Zubair, Lin, Nourbakhsh, and Palacios]{mcvay2020impact}
McVay,~E.; Zubair,~A.; Lin,~Y.; Nourbakhsh,~A.; Palacios,~T. Impact of Al2O3 passivation on the photovoltaic performance of vertical WSe2 Schottky junction solar cells. \emph{ACS Appl. Mater. Interfaces} \textbf{2020}, \emph{12}, 57987--57995\relax
\mciteBstWouldAddEndPuncttrue
\mciteSetBstMidEndSepPunct{\mcitedefaultmidpunct}
{\mcitedefaultendpunct}{\mcitedefaultseppunct}\relax
\EndOfBibitem
\bibitem[Islam \latin{et~al.}(2022)Islam, Ismael, Luthy, Kizilkaya, and Escarra]{Islam22}
Islam,~K.~M.; Ismael,~T.; Luthy,~C.; Kizilkaya,~O.; Escarra,~M.~D. Large-Area, High-Specific-Power Schottky-Junction Photovoltaics from CVD-Grown Monolayer MoS2. \emph{ACS Applied Materials \& Interfaces} \textbf{2022}, \emph{14}, 24281--24289\relax
\mciteBstWouldAddEndPuncttrue
\mciteSetBstMidEndSepPunct{\mcitedefaultmidpunct}
{\mcitedefaultendpunct}{\mcitedefaultseppunct}\relax
\EndOfBibitem
\bibitem[Liu \latin{et~al.}(2018)Liu, Guo, Zhu, Liao, Lee, Ding, Shakir, Gambin, Huang, and Duan]{liu2018approaching}
Liu,~Y.; Guo,~J.; Zhu,~E.; Liao,~L.; Lee,~S.-J.; Ding,~M.; Shakir,~I.; Gambin,~V.; Huang,~Y.; Duan,~X. Approaching the Schottky--Mott limit in van der Waals metal--semiconductor junctions. \emph{Nature} \textbf{2018}, \emph{557}, 696--700\relax
\mciteBstWouldAddEndPuncttrue
\mciteSetBstMidEndSepPunct{\mcitedefaultmidpunct}
{\mcitedefaultendpunct}{\mcitedefaultseppunct}\relax
\EndOfBibitem
\bibitem[Lee \latin{et~al.}(2014)Lee, Lee, van~der Zande, Chen, Li, Han, Cui, Arefe, Nuckolls, Heinz, Guo, Hone, and Kim]{lee2014atomically}
Lee,~C.-H.; Lee,~G.-H.; van~der Zande,~A.~M.; Chen,~W.; Li,~Y.; Han,~M.; Cui,~X.; Arefe,~G.; Nuckolls,~C.; Heinz,~T.~F.; Guo,~J.; Hone,~J.; Kim,~P. Atomically thin p--n junctions with van der Waals heterointerfaces. \emph{Nat. Nanotechnol.} \textbf{2014}, \emph{9}, 676--681\relax
\mciteBstWouldAddEndPuncttrue
\mciteSetBstMidEndSepPunct{\mcitedefaultmidpunct}
{\mcitedefaultendpunct}{\mcitedefaultseppunct}\relax
\EndOfBibitem
\bibitem[Memaran \latin{et~al.}(2015)Memaran, Pradhan, Lu, Rhodes, Ludwig, Zhou, Ogunsolu, Ajayan, Smirnov, Fernández-Domínguez, García-Vidal, and Balicas]{memaran2015pronounced}
Memaran,~S.; Pradhan,~N.~R.; Lu,~Z.; Rhodes,~D.; Ludwig,~J.; Zhou,~Q.; Ogunsolu,~O.; Ajayan,~P.~M.; Smirnov,~D.; Fernández-Domínguez,~A.~I.; García-Vidal,~F.~J.; Balicas,~L. Pronounced photovoltaic response from multilayered transition-metal dichalcogenides PN-junctions. \emph{Nano Lett.} \textbf{2015}, \emph{15}, 7532--7538\relax
\mciteBstWouldAddEndPuncttrue
\mciteSetBstMidEndSepPunct{\mcitedefaultmidpunct}
{\mcitedefaultendpunct}{\mcitedefaultseppunct}\relax
\EndOfBibitem
\bibitem[Cho \latin{et~al.}(2018)Cho, Song, Kang, and Kwon]{Cho18}
Cho,~A.-J.; Song,~M.-K.; Kang,~D.-W.; Kwon,~J.-Y. Two-Dimensional WSe2/MoS2 p–n Heterojunction-Based Transparent Photovoltaic Cell and Its Performance Enhancement by Fluoropolymer Passivation. \emph{ACS Appl. Mater. Interfaces} \textbf{2018}, \emph{10}, 35972--35977\relax
\mciteBstWouldAddEndPuncttrue
\mciteSetBstMidEndSepPunct{\mcitedefaultmidpunct}
{\mcitedefaultendpunct}{\mcitedefaultseppunct}\relax
\EndOfBibitem
\bibitem[Nassiri~Nazif \latin{et~al.}(2021)Nassiri~Nazif, Daus, Hong, Lee, Vaziri, Kumar, Nitta, Chen, Kananian, Islam, Kim, Park, Poon, Brongersma, Pop, and Saraswat]{Nazif21}
Nassiri~Nazif,~K.; Daus,~A.; Hong,~J.; Lee,~N.; Vaziri,~S.; Kumar,~A.; Nitta,~F.; Chen,~M.~E.; Kananian,~S.; Islam,~R.; Kim,~K.-H.; Park,~J.-H.; Poon,~A. S.~Y.; Brongersma,~M.~L.; Pop,~E.; Saraswat,~K.~C. High-specific-power flexible transition metal dichalcogenide solar cells. \emph{Nat. Commun.} \textbf{2021}, \emph{12}, 7034\relax
\mciteBstWouldAddEndPuncttrue
\mciteSetBstMidEndSepPunct{\mcitedefaultmidpunct}
{\mcitedefaultendpunct}{\mcitedefaultseppunct}\relax
\EndOfBibitem
\bibitem[Oz{\'o}rio \latin{et~al.}(2022)Oz{\'o}rio, Dias, Silveira, and Da~Silva]{ozorio2022theoretical}
Oz{\'o}rio,~M.~S.; Dias,~A.~C.; Silveira,~J.~F.; Da~Silva,~J.~L. Theoretical Investigation of the Role of Anion and Trivalent Cation Substitution in the Physical Properties of Lead-Free Zero-Dimensional Perovskites. \emph{J. Phys. Chem. C} \textbf{2022}, \emph{126}, 7245--7255\relax
\mciteBstWouldAddEndPuncttrue
\mciteSetBstMidEndSepPunct{\mcitedefaultmidpunct}
{\mcitedefaultendpunct}{\mcitedefaultseppunct}\relax
\EndOfBibitem
\bibitem[Hohenberg and Kohn(1964)Hohenberg, and Kohn]{HohenbergKohn1964}
Hohenberg,~P.; Kohn,~W. Inhomogeneous electron gas. \emph{Phys. Rev.} \textbf{1964}, \emph{136}, B864\relax
\mciteBstWouldAddEndPuncttrue
\mciteSetBstMidEndSepPunct{\mcitedefaultmidpunct}
{\mcitedefaultendpunct}{\mcitedefaultseppunct}\relax
\EndOfBibitem
\bibitem[Kohn and Sham(1965)Kohn, and Sham]{KohnSham1965}
Kohn,~W.; Sham,~L.~J. Self-consistent equations including exchange and correlation effects. \emph{Phys. Rev.} \textbf{1965}, \emph{140}, A1133\relax
\mciteBstWouldAddEndPuncttrue
\mciteSetBstMidEndSepPunct{\mcitedefaultmidpunct}
{\mcitedefaultendpunct}{\mcitedefaultseppunct}\relax
\EndOfBibitem
\bibitem[Giannozzi \latin{et~al.}(2009)Giannozzi, Baroni, Bonini, Calandra, Car, Cavazzoni, Ceresoli, Chiarotti, Cococcioni, Dabo, Corso, de~Gironcoli, Fabris, Fratesi, Gebauer, Gerstmann, Gougoussis, Kokalj, Lazzeri, Martin-Samos, Marzari, Mauri, Mazzarello, Paolini, Pasquarello, Paulatto, Sbraccia, Scandolo, Sclauzero, Seitsonen, Smogunov, Umari, and Wentzcovitch]{qe}
Giannozzi,~P.; Baroni,~S.; Bonini,~N.; Calandra,~M.; Car,~R.; Cavazzoni,~C.; Ceresoli,~D.; Chiarotti,~G.~L.; Cococcioni,~M.; Dabo,~I.; Corso,~A.~D.; de~Gironcoli,~S.; Fabris,~S.; Fratesi,~G.; Gebauer,~R.; Gerstmann,~U.; Gougoussis,~C.; Kokalj,~A.; Lazzeri,~M.; Martin-Samos,~L.; Marzari,~N.; Mauri,~F.; Mazzarello,~R.; Paolini,~S.; Pasquarello,~A.; Paulatto,~L.; Sbraccia,~C.; Scandolo,~S.; Sclauzero,~G.; Seitsonen,~A.~P.; Smogunov,~A.; Umari,~P.; Wentzcovitch,~R.~M. QUANTUM ESPRESSO: a modular and open-source software project for quantum simulations of materials. \emph{J. Phys.:~Condens. Matter} \textbf{2009}, \emph{21}, 395502\relax
\mciteBstWouldAddEndPuncttrue
\mciteSetBstMidEndSepPunct{\mcitedefaultmidpunct}
{\mcitedefaultendpunct}{\mcitedefaultseppunct}\relax
\EndOfBibitem
\bibitem[Perdew \latin{et~al.}(1996)Perdew, Burke, and Ernzerhof]{pbe}
Perdew,~J.~P.; Burke,~K.; Ernzerhof,~M. Generalized gradient approximation made simple. \emph{Phys. Rev. Lett.} \textbf{1996}, \emph{77}, 3865\relax
\mciteBstWouldAddEndPuncttrue
\mciteSetBstMidEndSepPunct{\mcitedefaultmidpunct}
{\mcitedefaultendpunct}{\mcitedefaultseppunct}\relax
\EndOfBibitem
\bibitem[Hamann(2013)]{ONCVPSP}
Hamann,~D.~R. Optimized norm-conserving Vanderbilt pseudopotentials. \emph{Phys. Rev. B} \textbf{2013}, \emph{88}, 085117\relax
\mciteBstWouldAddEndPuncttrue
\mciteSetBstMidEndSepPunct{\mcitedefaultmidpunct}
{\mcitedefaultendpunct}{\mcitedefaultseppunct}\relax
\EndOfBibitem
\bibitem[Grimme(2006)]{pbe-d2}
Grimme,~S. Semiempirical GGA-type density functional constructed with a long-range dispersion correction. \emph{J. Comput. Chem.} \textbf{2006}, \emph{27}, 1787--1799\relax
\mciteBstWouldAddEndPuncttrue
\mciteSetBstMidEndSepPunct{\mcitedefaultmidpunct}
{\mcitedefaultendpunct}{\mcitedefaultseppunct}\relax
\EndOfBibitem
\bibitem[Hedin(1965)]{hedin1965}
Hedin,~L. New Method for Calculating the One-Particle Green's Function with Application to the Electron-Gas Problem. \emph{Phys. Rev.} \textbf{1965}, \emph{139}, A796--A823\relax
\mciteBstWouldAddEndPuncttrue
\mciteSetBstMidEndSepPunct{\mcitedefaultmidpunct}
{\mcitedefaultendpunct}{\mcitedefaultseppunct}\relax
\EndOfBibitem
\bibitem[Marini \latin{et~al.}(2009)Marini, Hogan, Grüning, and Varsano]{yambo}
Marini,~A.; Hogan,~C.; Grüning,~M.; Varsano,~D. yambo: An ab initio tool for excited state calculations. \emph{Comput. Phys. Commun.} \textbf{2009}, \emph{180}, 1392--1403\relax
\mciteBstWouldAddEndPuncttrue
\mciteSetBstMidEndSepPunct{\mcitedefaultmidpunct}
{\mcitedefaultendpunct}{\mcitedefaultseppunct}\relax
\EndOfBibitem
\bibitem[Bruneval and Gonze(2008)Bruneval, and Gonze]{BGterminatorPRB2008}
Bruneval,~F.; Gonze,~X. Accurate $GW$ self-energies in a plane-wave basis using only a few empty states: Towards large systems. \emph{Phys. Rev. B} \textbf{2008}, \emph{78}, 085125\relax
\mciteBstWouldAddEndPuncttrue
\mciteSetBstMidEndSepPunct{\mcitedefaultmidpunct}
{\mcitedefaultendpunct}{\mcitedefaultseppunct}\relax
\EndOfBibitem
\bibitem[Rohlfing and Louie(2000)Rohlfing, and Louie]{louie2000PRB}
Rohlfing,~M.; Louie,~S.~G. Electron-hole excitations and optical spectra from first principles. \emph{Phys. Rev. B} \textbf{2000}, \emph{62}, 4927--4944\relax
\mciteBstWouldAddEndPuncttrue
\mciteSetBstMidEndSepPunct{\mcitedefaultmidpunct}
{\mcitedefaultendpunct}{\mcitedefaultseppunct}\relax
\EndOfBibitem
\bibitem[Kammerlander \latin{et~al.}(2012)Kammerlander, Botti, Marques, Marini, and Attaccalite]{kammerlander2012}
Kammerlander,~D.; Botti,~S.; Marques,~M.~A.; Marini,~A.; Attaccalite,~C. Speeding up the solution of the Bethe-Salpeter equation by a double-grid method and Wannier interpolation. \emph{Phys. Rev. B} \textbf{2012}, \emph{86}, 125203\relax
\mciteBstWouldAddEndPuncttrue
\mciteSetBstMidEndSepPunct{\mcitedefaultmidpunct}
{\mcitedefaultendpunct}{\mcitedefaultseppunct}\relax
\EndOfBibitem
\bibitem[SI()]{SI}
Supporting information: convergence analyses of $GW$ and BSE calculations, and results of photovoltaic properties of the bulk TMDCs within Shockley-Queisser formalism.\relax
\mciteBstWouldAddEndPunctfalse
\mciteSetBstMidEndSepPunct{\mcitedefaultmidpunct}
{}{\mcitedefaultseppunct}\relax
\EndOfBibitem
\bibitem[Wildervanck and Jellinek(1964)Wildervanck, and Jellinek]{wildervanck1964MoS2}
Wildervanck,~J.; Jellinek,~F. Preparation and crystallinity of molybdenum and tungsten sulfides. \emph{Z. Anorg. Allg. Chem.} \textbf{1964}, \emph{328}, 309\relax
\mciteBstWouldAddEndPuncttrue
\mciteSetBstMidEndSepPunct{\mcitedefaultmidpunct}
{\mcitedefaultendpunct}{\mcitedefaultseppunct}\relax
\EndOfBibitem
\bibitem[Schutte \latin{et~al.}(1987)Schutte, {De Boer}, and Jellinek]{Schutte1987WS2-WSe2}
Schutte,~W.; {De Boer},~J.; Jellinek,~F. Crystal structures of tungsten disulfide and diselenide. \emph{J. Solid State Chem.} \textbf{1987}, \emph{70}, 207--209\relax
\mciteBstWouldAddEndPuncttrue
\mciteSetBstMidEndSepPunct{\mcitedefaultmidpunct}
{\mcitedefaultendpunct}{\mcitedefaultseppunct}\relax
\EndOfBibitem
\bibitem[Evans and Hazelwood(1971)Evans, and Hazelwood]{evans1971MoSe2}
Evans,~B.; Hazelwood,~R. Optical and structural properties of MoSe2. \emph{Phys. Status Solidi (A)} \textbf{1971}, \emph{4}, 181\relax
\mciteBstWouldAddEndPuncttrue
\mciteSetBstMidEndSepPunct{\mcitedefaultmidpunct}
{\mcitedefaultendpunct}{\mcitedefaultseppunct}\relax
\EndOfBibitem
\bibitem[Kim \latin{et~al.}(2016)Kim, Rhim, Kim, Kim, and Park]{kim2016determination}
Kim,~B.~S.; Rhim,~J.-W.; Kim,~B.; Kim,~C.; Park,~S.~R. Determination of the band parameters of bulk 2H-MX2 (M= Mo, W; X= S, Se) by angle-resolved photoemission spectroscopy. \emph{Sci. Rep.} \textbf{2016}, \emph{6}, 36389\relax
\mciteBstWouldAddEndPuncttrue
\mciteSetBstMidEndSepPunct{\mcitedefaultmidpunct}
{\mcitedefaultendpunct}{\mcitedefaultseppunct}\relax
\EndOfBibitem
\bibitem[Beal \latin{et~al.}(1976)Beal, Liang, and Hughes]{beal1976}
Beal,~A.; Liang,~W.; Hughes,~H. Kramers-Kronig analysis of the reflectivity spectra of 3R-WS2 and 2H-WSe2. \emph{J. Phys. C: Solid State Phys.} \textbf{1976}, \emph{9}, 2449\relax
\mciteBstWouldAddEndPuncttrue
\mciteSetBstMidEndSepPunct{\mcitedefaultmidpunct}
{\mcitedefaultendpunct}{\mcitedefaultseppunct}\relax
\EndOfBibitem
\bibitem[Beal and Hughes(1979)Beal, and Hughes]{beal1979}
Beal,~A.; Hughes,~H. Kramers-Kronig analysis of the reflectivity spectra of 2H-MoS2, 2H-MoSe2 and 2H-MoTe2. \emph{J. Phys. C: Solid State Phys.} \textbf{1979}, \emph{12}, 881\relax
\mciteBstWouldAddEndPuncttrue
\mciteSetBstMidEndSepPunct{\mcitedefaultmidpunct}
{\mcitedefaultendpunct}{\mcitedefaultseppunct}\relax
\EndOfBibitem
\bibitem[Li \latin{et~al.}(2014)Li, Chernikov, Zhang, Rigosi, Hill, Van Der~Zande, Chenet, Shih, Hone, and Heinz]{li2014measurement}
Li,~Y.; Chernikov,~A.; Zhang,~X.; Rigosi,~A.; Hill,~H.~M.; Van Der~Zande,~A.~M.; Chenet,~D.~A.; Shih,~E.-M.; Hone,~J.; Heinz,~T.~F. Measurement of the optical dielectric function of monolayer transition-metal dichalcogenides: MoS 2, Mo S e 2, WS 2, and WS e 2. \emph{Phys. Rev. B} \textbf{2014}, \emph{90}, 205422\relax
\mciteBstWouldAddEndPuncttrue
\mciteSetBstMidEndSepPunct{\mcitedefaultmidpunct}
{\mcitedefaultendpunct}{\mcitedefaultseppunct}\relax
\EndOfBibitem
\bibitem[Wang \latin{et~al.}(2018)Wang, Chernikov, Glazov, Heinz, Marie, Amand, and Urbaszek]{wang2018colloquium}
Wang,~G.; Chernikov,~A.; Glazov,~M.~M.; Heinz,~T.~F.; Marie,~X.; Amand,~T.; Urbaszek,~B. Colloquium: Excitons in atomically thin transition metal dichalcogenides. \emph{Rev. Mod. Phys.} \textbf{2018}, \emph{90}, 021001\relax
\mciteBstWouldAddEndPuncttrue
\mciteSetBstMidEndSepPunct{\mcitedefaultmidpunct}
{\mcitedefaultendpunct}{\mcitedefaultseppunct}\relax
\EndOfBibitem
\bibitem[Yu and Zunger(2012)Yu, and Zunger]{Yu-Zunger-PRL-SLME}
Yu,~L.; Zunger,~A. Identification of Potential Photovoltaic Absorbers Based on First-Principles Spectroscopic Screening of Materials. \emph{Phys. Rev. Lett.} \textbf{2012}, \emph{108}, 068701\relax
\mciteBstWouldAddEndPuncttrue
\mciteSetBstMidEndSepPunct{\mcitedefaultmidpunct}
{\mcitedefaultendpunct}{\mcitedefaultseppunct}\relax
\EndOfBibitem
\bibitem[Shockley and Queisser(1961)Shockley, and Queisser]{sq-limit}
Shockley,~W.; Queisser,~H.~J. Detailed Balance Limit of Efficiency of p‐n Junction Solar Cells. \emph{J. Appl. Phys.} \textbf{1961}, \emph{32}, 510--519\relax
\mciteBstWouldAddEndPuncttrue
\mciteSetBstMidEndSepPunct{\mcitedefaultmidpunct}
{\mcitedefaultendpunct}{\mcitedefaultseppunct}\relax
\EndOfBibitem
\bibitem[Mohamed \latin{et~al.}(2017)Mohamed, Lim, Wang, Koirala, Mouri, Shinokita, Miyauchi, and Matsuda]{mohamed2017}
Mohamed,~N.~B.; Lim,~H.~E.; Wang,~F.; Koirala,~S.; Mouri,~S.; Shinokita,~K.; Miyauchi,~Y.; Matsuda,~K. Long radiative lifetimes of excitons in monolayer transition-metal dichalcogenides MX2 (M= Mo, W; X= S, Se). \emph{Appl. Phys. Express} \textbf{2017}, \emph{11}, 015201\relax
\mciteBstWouldAddEndPuncttrue
\mciteSetBstMidEndSepPunct{\mcitedefaultmidpunct}
{\mcitedefaultendpunct}{\mcitedefaultseppunct}\relax
\EndOfBibitem
\bibitem[Bercx \latin{et~al.}(2016)Bercx, Sarmadian, Saniz, Partoens, and Lamoen]{bercx2016first}
Bercx,~M.; Sarmadian,~N.; Saniz,~R.; Partoens,~B.; Lamoen,~D. First-principles analysis of the spectroscopic limited maximum efficiency of photovoltaic absorber layers for CuAu-like chalcogenides and silicon. \emph{Phys. Chem. Chem. Phys.} \textbf{2016}, \emph{18}, 20542--20549\relax
\mciteBstWouldAddEndPuncttrue
\mciteSetBstMidEndSepPunct{\mcitedefaultmidpunct}
{\mcitedefaultendpunct}{\mcitedefaultseppunct}\relax
\EndOfBibitem
\bibitem[Moujaes and Dias(2023)Moujaes, and Dias]{moujaes2023}
Moujaes,~E.~A.; Dias,~A.~C. On the excitonic effects of the 1T and 1OT phases of PdS2, PdSe2, and PdSSe monolayers. \emph{J. Phys. Chem. Solids} \textbf{2023}, \emph{182}, 111573\relax
\mciteBstWouldAddEndPuncttrue
\mciteSetBstMidEndSepPunct{\mcitedefaultmidpunct}
{\mcitedefaultendpunct}{\mcitedefaultseppunct}\relax
\EndOfBibitem
\bibitem[Dreessen \latin{et~al.}(2023)Dreessen, Zanoni, Gil-Escrig, Rodkey, Khan, Laquai, Sessolo, Roldán-Carmona, and Bolink]{Dreessen23}
Dreessen,~C.; Zanoni,~K. P.~S.; Gil-Escrig,~L.; Rodkey,~N.; Khan,~J.~I.; Laquai,~F.; Sessolo,~M.; Roldán-Carmona,~C.; Bolink,~H.~J. When JV Curves Conceal Material Improvements: The Relevance of Photoluminescence Measurements in the Optimization of Perovskite Solar Cells. \emph{Adv. Opt. Mater.} \textbf{2023}, \emph{n/a}, 2301019\relax
\mciteBstWouldAddEndPuncttrue
\mciteSetBstMidEndSepPunct{\mcitedefaultmidpunct}
{\mcitedefaultendpunct}{\mcitedefaultseppunct}\relax
\EndOfBibitem
\bibitem[Jariwala \latin{et~al.}(2017)Jariwala, Davoyan, Wong, and Atwater]{jariwala2017van}
Jariwala,~D.; Davoyan,~A.~R.; Wong,~J.; Atwater,~H.~A. Van der Waals materials for atomically-thin photovoltaics: promise and outlook. \emph{ACS Photonics} \textbf{2017}, \emph{4}, 2962--2970\relax
\mciteBstWouldAddEndPuncttrue
\mciteSetBstMidEndSepPunct{\mcitedefaultmidpunct}
{\mcitedefaultendpunct}{\mcitedefaultseppunct}\relax
\EndOfBibitem
\bibitem[Amani \latin{et~al.}(2015)Amani, Lien, Kiriya, Xiao, Azcatl, Noh, Madhvapathy, Addou, KC, Dubey, Cho, Wallace, Lee, He, Ager, Zhang, Yablonovitch, and Javey]{amani2015near}
Amani,~M.; Lien,~D.-H.; Kiriya,~D.; Xiao,~J.; Azcatl,~A.; Noh,~J.; Madhvapathy,~S.~R.; Addou,~R.; KC,~S.; Dubey,~M.; Cho,~K.; Wallace,~R.~M.; Lee,~S.-C.; He,~J.-H.; Ager,~J.~W.; Zhang,~X.; Yablonovitch,~E.; Javey,~A. Near-unity photoluminescence quantum yield in MoS2. \emph{Science} \textbf{2015}, \emph{350}, 1065--1068\relax
\mciteBstWouldAddEndPuncttrue
\mciteSetBstMidEndSepPunct{\mcitedefaultmidpunct}
{\mcitedefaultendpunct}{\mcitedefaultseppunct}\relax
\EndOfBibitem
\bibitem[Amani \latin{et~al.}(2016)Amani, Taheri, Addou, Ahn, Kiriya, Lien, Ager~III, Wallace, and Javey]{amani2016recombination}
Amani,~M.; Taheri,~P.; Addou,~R.; Ahn,~G.~H.; Kiriya,~D.; Lien,~D.-H.; Ager~III,~J.~W.; Wallace,~R.~M.; Javey,~A. Recombination kinetics and effects of superacid treatment in sulfur-and selenium-based transition metal dichalcogenides. \emph{Nano Lett.} \textbf{2016}, \emph{16}, 2786--2791\relax
\mciteBstWouldAddEndPuncttrue
\mciteSetBstMidEndSepPunct{\mcitedefaultmidpunct}
{\mcitedefaultendpunct}{\mcitedefaultseppunct}\relax
\EndOfBibitem
\bibitem[Mak \latin{et~al.}(2010)Mak, Lee, Hone, Shan, and Heinz]{mak2010atomically}
Mak,~K.~F.; Lee,~C.; Hone,~J.; Shan,~J.; Heinz,~T.~F. Atomically thin MoS 2: a new direct-gap semiconductor. \emph{Phys. Rev. Lett.} \textbf{2010}, \emph{105}, 136805\relax
\mciteBstWouldAddEndPuncttrue
\mciteSetBstMidEndSepPunct{\mcitedefaultmidpunct}
{\mcitedefaultendpunct}{\mcitedefaultseppunct}\relax
\EndOfBibitem
\bibitem[Ozdemir and Barone(2020)Ozdemir, and Barone]{ozdemir2020thickness}
Ozdemir,~B.; Barone,~V. Thickness dependence of solar cell efficiency in transition metal dichalcogenides MX2 (M: Mo, W; X: S, Se, Te). \emph{Sol. Energy Mater Sol. Cells} \textbf{2020}, \emph{212}, 110557\relax
\mciteBstWouldAddEndPuncttrue
\mciteSetBstMidEndSepPunct{\mcitedefaultmidpunct}
{\mcitedefaultendpunct}{\mcitedefaultseppunct}\relax
\EndOfBibitem
\bibitem[Choudhary \latin{et~al.}(2019)Choudhary, Bercx, Jiang, Pachter, Lamoen, and Tavazza]{choudhary2019accelerated}
Choudhary,~K.; Bercx,~M.; Jiang,~J.; Pachter,~R.; Lamoen,~D.; Tavazza,~F. Accelerated discovery of efficient solar cell materials using quantum and machine-learning methods. \emph{Chem. Mater.} \textbf{2019}, \emph{31}, 5900--5908\relax
\mciteBstWouldAddEndPuncttrue
\mciteSetBstMidEndSepPunct{\mcitedefaultmidpunct}
{\mcitedefaultendpunct}{\mcitedefaultseppunct}\relax
\EndOfBibitem
\end{mcitethebibliography}
\clearpage
\begin{figure}
    \centering
    \includegraphics[width=.8\textwidth]{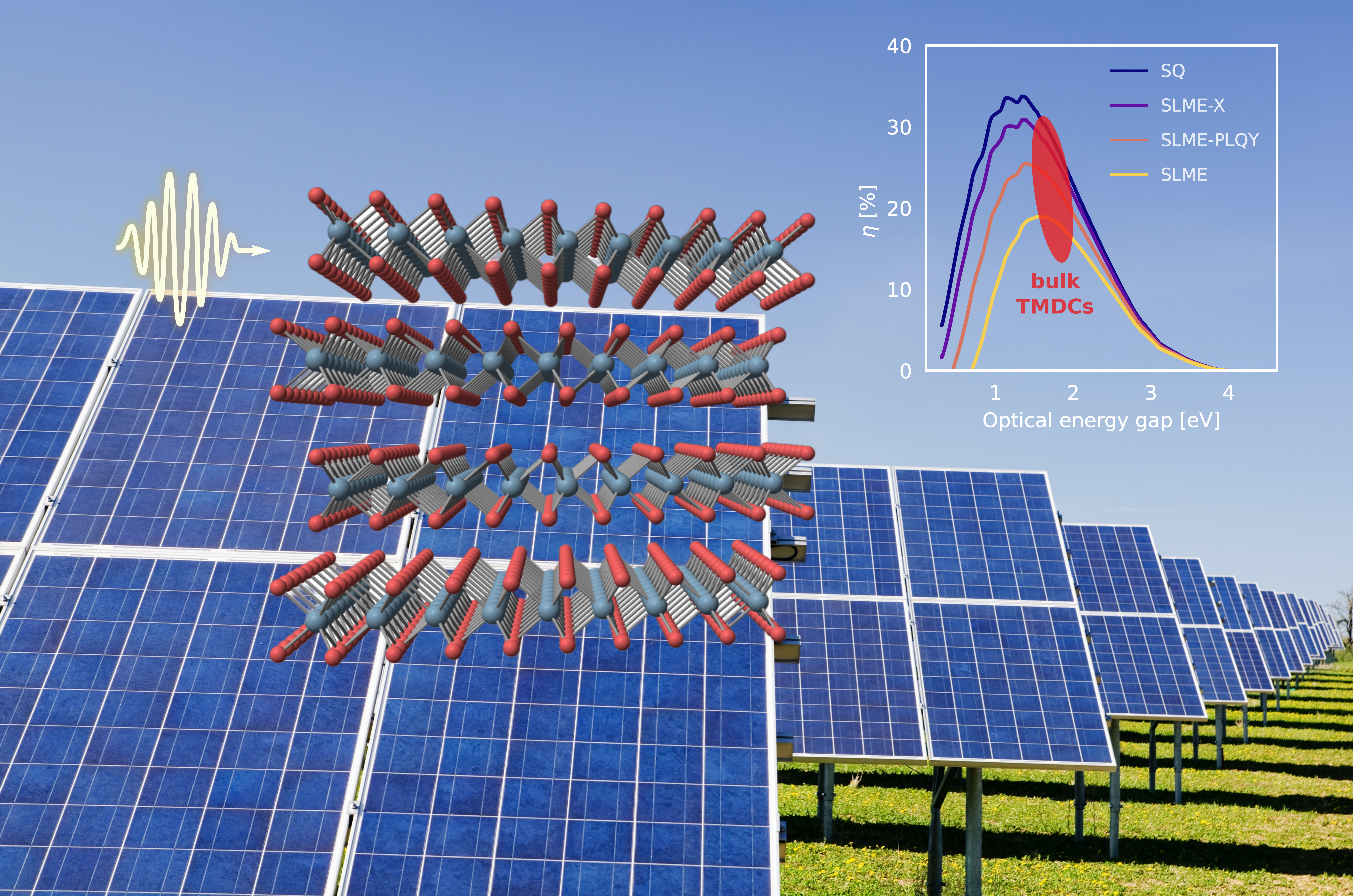}
    \caption*{\textbf{TOC Graphic}}
\end{figure}
\end{document}